\documentclass[12pt,a4paper]{article}

\usepackage[nosort]{cite}
\usepackage{bbm}
\usepackage{amsmath}
\usepackage{amssymb}
\usepackage{amscd}
\usepackage{slashed}
\usepackage{array}
\usepackage{rotating}
\usepackage{float}

\makeatletter \@addtoreset{equation}{section} \makeatother

\makeatletter
\let\old@startsection=\@startsection
\let\oldl@section=\l@section
\renewcommand{\@startsection}[6]{\old@startsection{#1}{#2}{#3}{#4}{#5}{#6\mathversion{bold}}}
\renewcommand{\l@section}[2]{\oldl@section{\mathversion{bold}#1}{#2}}
\makeatother

\makeatletter
\let\old@makecaption=\@makecaption
\def\@makecaption{\small\old@makecaption}
\makeatother

\renewcommand{\thefootnote}{\arabic{footnote}}
\let\oldPhi=\Phi
\let\oldPsi=\Psi
\let\oldGamma=\Gamma
\let\oldDelta=\Delta
\let\oldSigma=\Sigma
\let\oldTheta=\Theta
\let\oldPi=\Pi
\let\oldUpsilon=\Upsilon
\renewcommand{\Phi}{\mathnormal{\oldPhi}}
\renewcommand{\Psi}{\mathnormal{\oldPsi}}
\renewcommand{\Gamma}{\mathnormal{\oldGamma}}
\renewcommand{\Sigma}{\mathnormal{\oldSigma}}
\renewcommand{\Delta}{\mathnormal{\oldDelta}}
\renewcommand{\Theta}{\mathnormal{\oldTheta}}
\renewcommand{\Pi}{\mathnormal{\oldPi}}
\renewcommand{\Upsilon}{\mathnormal{\oldUpsilon}}

\newcommand{\superN}{\mathcal{N}}

\newcommand{\Lagr}{\mathcal{L}}

\newcommand{\tr}{\mathop{\mathrm{tr}}}

\newcommand{\diag}{\mathop{\mathrm{diag}}}

\newcommand{\Integers}{\mathbbm{Z}}
\newcommand{\Reals}{\mathbbm{R}}

\newcommand{\Sphere}{S}  
\newcommand{\AdS}{\mathrm{AdS}}

\ifx\genfrac\sdflkaj

\else

\fi
\newcommand{\sfrac}[2]{{\textstyle\frac{#1}{#2}}}
\newcommand{\half}{\sfrac{1}{2}}

\newcommand{\Half}{\frac{1}{2}}

\newcommand{\et}{\frac{\eps}{3}}
\newcommand{\eft}{\frac{4 \eps}{3}}
\newcommand{\ett}{\frac{2 \eps}{3}}
\newcommand{\evt}{\frac{5 \eps}{3}}

\newcommand{\rep}[1]{{\mathbf{#1}}}
\newcommand{\matr}[2]{\left(\begin{array}{#1}#2\end{array}\right)}

\newcommand{\grp}[1]{\mathrm{#1}}

\newcommand{\grU}{\grp{U}}
\newcommand{\grSU}{\grp{SU}}
\newcommand{\grSO}{\grp{SO}}

\newcommand{\grOsp}{\grp{Osp}}

\newcommand{\lrbrk}[1]{\left(#1\right)}
\newcommand{\bigbrk}[1]{\bigl(#1\bigr)}

\newcommand{\lrsbrk}[1]{\left[#1\right]}

\newcommand{\Bigsbrk}[1]{\Bigl[#1\Bigr]}

\newcommand{\abs}[1]{{|#1|}}

\newcommand{\lrfloor}[1]{\left\lfloor #1 \right\rfloor}

\newcommand{\nn}{\nonumber}

\newcommand{\nl}[1][0pt]{\nonumber\\[#1]&\hspace{-4\arraycolsep}&\mathord{}}

\newcommand{\earel}[1]{\mathrel{}&\hspace{-2\arraycolsep}#1\hspace{-2\arraycolsep}&\mathrel{}}
\newcommand{\eq}{\earel{=}}

\def\[{\begin{equation}}
\def\]{\end{equation}}

\newcommand{\be}{\begin{eqnarray}}
\newcommand{\ee}{\end{eqnarray}}

\makeatletter
\def\mr@ignsp#1 {\ifx\:#1\@empty\else #1\expandafter\mr@ignsp\fi}%
\newcommand{\multiref}[1]{\begingroup
\xdef\mr@no@sparg{\expandafter\mr@ignsp#1 \: }%
\def\mr@comma{}%
\@for\mr@refs:=\mr@no@sparg\do{\mr@comma\def\mr@comma{,}\ref{\mr@refs}}%
\endgroup}
\makeatother

\newcommand{\hypref}[2]{\ifx\href\asklfhas #2\else\href{#1}{#2}\fi}

\newcommand{\secref}[1]{Sec.~\multiref{#1}}

\newcommand{\appref}[1]{App.~\multiref{#1}}

\newcommand{\tabref}[1]{Tab.~\multiref{#1}}

\renewcommand{\eqref}[1]{(\multiref{#1})}

\ifx\href\asklfhas\newcommand{\href}[2]{#2}\fi

\newcommand{\comma}{\quad,\quad}

\newcommand{\levi}{\epsilon}

\newcommand{\eps}{\varepsilon}

\newcommand{\ZZ}{\mathcal{Z}}
\newcommand{\WW}{\mathcal{W}}

\newcommand{\VV}{\mathcal{V}}

\newcommand{\superW}{\mathrm{W}}

\newcommand{\ha}{{\hat{a}}}
\newcommand{\hb}{{\hat{b}}}
\newcommand{\hc}{{\hat{c}}}
\newcommand{\hd}{{\hat{d}}}

\begin{document}

\thispagestyle{empty}
\begin{flushright}\footnotesize
\texttt{arXiv:0809.3773}\\
\texttt{PUPT-2282}\vspace{10mm}
\end{flushright}

\renewcommand{\thefootnote}{\fnsymbol{footnote}}
\setcounter{footnote}{0}

\begin{center}
{\Large\textbf{\mathversion{bold}
AdS$_4$/CFT$_3$\\
Squashed, Stretched and Warped
}\par}

\vspace{1.5cm}

\textrm{Igor R. Klebanov$^{a,b}$, Thomas Klose$^{a,b}$ and Arvind Murugan$^{a}$} \vspace{8mm} \\
\textit{
$^a$Joseph Henry Laboratories and $^b$Princeton Center for Theoretical Science \\
Princeton University, Princeton, NJ 08544, USA
} \\
\texttt{klebanov,tklose,arvind@princeton.edu}

\par\vspace{14mm}

\textbf{Abstract} \vspace{5mm}

\begin{minipage}{14cm}

We use group theoretic methods to calculate the spectrum of short multiplets around the extremum of $\superN=8$ gauged supergravity potential which possesses $\superN=2$ supersymmetry and $\grSU(3)$ global symmetry. Upon uplifting to M-theory, it describes a warped product of $AdS_4$ and a certain squashed and stretched 7-sphere. We find quantum numbers in agreement with those of the gauge invariant operators in the $\superN=2$ superconformal Chern-Simons theory recently proposed to be the dual of this M-theory background. This theory is obtained from the $\grU(N)\times \grU(N)$ theory through deforming the superpotential by a term quadratic in one of the superfields. To construct this model explicitly, one needs to employ monopole operators whose complete understanding is still lacking. However, for the $\grU(2)\times\grU(2)$ gauge theory we make a proposal for the form of the monopole operators which has a number of desired properties. In particular, this proposal implies enhanced symmetry of the $\grU(2)\times\grU(2)$ ABJM theory for $k=1,2$; it makes its similarity to and subtle difference from the BLG theory quite explicit.

\end{minipage}

\end{center}

\vspace{1.5cm}

\newpage
\setcounter{page}{1}
\renewcommand{\thefootnote}{\arabic{footnote}}
\setcounter{footnote}{0}

\hrule
\tableofcontents
\vspace{8mm}
\hrule
\vspace{4mm}

\setlength{\extrarowheight}{5pt}

\section{Introduction and Summary}

During the recent months, remarkable progress has taken place in understanding the world volume theory of coincident M2-branes. This was precipitated by the discovery by Bagger and Lambert \cite{Bagger:2006sk, Bagger:2007jr, Bagger:2007vi}, and by Gustavsson \cite{Gustavsson:2007vu}, of the $2+1$ dimensional superconformal Chern-Simons theory with the maximal $\superN=8$ supersymmetry and manifest $\grSO(8)$ R-symmetry (these papers were inspired in part by the ideas of \cite{Basu:2004ed,Schwarz:2004yj}). The Bagger-Lambert-Gustavsson (BLG) 3-algebra construction was, under the assumption of manifest unitarity, limited to the gauge group $\grSO(4)$. This BLG theory is conveniently reformulated as an $\grSU(2)\times \grSU(2)$ gauge theory with conventional Chern-Simons terms having opposite levels $k$ and $-k$ \cite{VanRaamsdonk:2008ft,Bandres:2008vf}. While the extension to more general gauge groups at first appeared to be difficult, major progress was eventually achieved by Aharony, Bergman, Jafferis and Maldacena (ABJM) \cite{Aharony:2008ug} who proposed a $\grU(N)\times \grU(N)$ Chern-Simons gauge theory with levels $k$ and $-k$ as a dual description of $N$ M2-branes placed at the singularity of $\Reals^8/\Integers_k$. The $\Integers_k$ acts by simultaneous rotation in the four planes; for $k>2$ this orbifold preserves only $\superN=6$ supersymmetry. ABJM gave strong evidence that their Chern-Simons gauge theory indeed possesses this amount of supersymmetry, and further work in \cite{Benna:2008zy,Bandres:2008ry} provided confirmation of this claim. Furthermore, for $k=1,2$ the supersymmetry of the orbifold, and therefore of the gauge theory, is expected to be enhanced to $\superN=8$. This is not manifest in the ABJM theory. Generally, inclusion of monopole operators is expected to play a crucial role both in the enhancement of the supersymmetry and in describing the full spectrum of gauge invariant operators. Explicit construction of these monopoles in ABJM theory was initiated in \cite{Berenstein:2008dc} but not all properties required here have been established. We will make some comments on these monopole operators, although our explicit calculations will mostly refer to the $\grU(2)\times\grU(2)$ case. Without the use of monopole operators one can make at most $\superN=6$ supersymmetry manifest in theories with higher rank gauge groups. These theories were classified in \cite{Schnabl:2008wj}.

The explicit formulation of highly supersymmetric theories on M2-branes raises hope that one can also formulate AdS$_4$/CFT$_3$ dualities with reduced supersymmetry. To this end one may consider orbifolds or orientifolds of the BLG and ABJM theories \cite{Fuji:2008yj,Hosomichi:2008jd,Benna:2008zy,Imamura:2008nn,Terashima:2008ba,Hosomichi:2008jb,Aharony:2008gk}. But it is also interesting to look for gauge theories that are dual to backgrounds that do not locally look like $AdS_4\times S^7$. Recent steps in this direction were made in \cite{Ooguri:2008dk} where a dual to the $\superN=1$ supersymmetric squashing of the $S^7$ was proposed, and in \cite{Martelli:2008rt,Martelli:2008si,Hanany:2008cd,Ueda:2008hx,Imamura:2008qs,Hanany:2008fj,Franco:2008um} where $S^7$ was replaced by manifolds preserving $\superN=2$ or $\superN=3$ supersymmetry. In the present paper we continue the program begun in \cite{Benna:2008zy} (see also \cite{Ahn:2008ya}) where an $\superN=2$ superpotential deformation of $k=1,2$ ABJM theory by a term quadratic in one of the bi-fundamental superfields was shown to create an RG flow leading to a new Chern-Simons CFT with $\superN=2$ supersymmetry and $\grSU(3)$ global symmetry. This CFT was conjectured to be dual to Warner's $\grSU(3)\times \grU(1)_R$ invariant extremum \cite{Warner:1983vz} of the potential in the gauged $\superN=8$ supergravity \cite{deWit:1982ig}. This extremum was uplifted to an 11-dimensional warped $AdS_4$ background containing a `squashed and stretched' 7-sphere \cite{Corrado:2001nv} (this terminology suggested the title of our paper). This background is of the Englert type in that there is a 4-form field strength turned on in the 7-sphere directions \cite{Englert:1982vs}. As a result, it breaks parity (reflection of one world volume direction accompanied by $C_{IJK}\rightarrow - C_{IJK}$) and we will show that the parity is also broken in the gauge theory. The $\superN=2$ superconformal symmetry of this background facilitates the comparison, via the AdS/CFT map \cite{Maldacena:1997re,Gubser:1998bc,Witten:1998qj}, of the $\grSU(3)\times \grU(1)_R$ quantum numbers and energies of supergravity fluctuations with those of the gauge invariant operators in the Chern-Simons CFT. One interesting feature of the gauge theory is that far in the IR the effective superpotential is sextic in the bi-fundamental chiral superfields. The marginality then requires that their $\grU(1)_R$ charges equal $1/3$.

On the supergravity side, the analysis of the $\grSU(3)\times \grU(1)_R$ quantum numbers was initiated in \cite{Nicolai:1985hs}, where some low-lying supermultiplets were constructed. It was noted that $\superN=2$ supersymmetry allowed for two alternative ways of assigning $\grSU(3)\times \grU(1)_R$ quantum numbers; however, the two $\grU(1)$ embeddings were found to be essentially equivalent at the lowest level \cite{Nicolai:1985hs}. Indeed, in \appref{app:dressing} we will show that there is no difference between the two choices in the values of $m^2$ in $AdS_4$ corresponding to the lowest hypermultiplet studied in \cite{Nicolai:1985hs}. The only difference concerns the choice of branches in the square root formula entering the operator dimensions. However, working only at the level of superconformal symmetry alone and not doing an explicit KK reduction, we show that these two choices of assigning $\grSU(3)\times \grU(1)_R$ quantum numbers lead to completely different spectra at higher levels. It should be stressed that even though such a group theoretical method does not necessarily lead to a unique answer, it is a rather efficient tool to gain insights into the spectrum. The first assignment of charges, which we will call Scenario I, produces agreement with the proposed gauge theory. The second one, Scenario II, which for the lowest hypermultiplet was spelled out in \cite{Nicolai:1985hs}, turns out not to agree with our gauge theory proposal. In general, the mass spectra resulting from the two scenarios are distinct and hence an explicit KK reduction could agree with only one of them. We show that when considering higher massive multiplets, Scenario II does not appear to give a spectrum characteristic of KK reduction.\footnote{In fact, some evidence for the correctness of scenario I has recently been obtained also from solving the minimally coupled scalar equation in the background under consideration \cite{KPR}.}

In \secref{sec:sugra} we review the gauged supergravity analysis of multiplets from \cite{Nicolai:1985hs}, extend this work to higher levels and scrutinize the differences between Scenarios I and II for grouping fluctuations into supermultiplets. In \secref{sec:GaugeTheory} we review the ABJM theory and its relevant deformation, emphasizing the important role of monopole operators. We show how the expected symmetries of the theory emerge for $N=2$ for a certain form of these operators. In \secref{sec:Matching} we analyze the short multiplets of chiral operators, demonstrating agreement with the gauged supergravity. The general structure of $\superN=2$ supermultiplets, and their specific examples occurring in this theory, as well as some comments on the monopole operators are left for the Appendices.

\section{Supergravity side}
\label{sec:sugra}

The supergravity background proposed in \cite{Benna:2008zy} as a dual to the mass deformed ABJM gauge theory (a related yet somewhat different proposal independently appeared in \cite{Ahn:2008ya}) was first found by Warner \cite{Warner:1983vz} as one of several non-trivial extrema of the gauged $\superN=8$ SUGRA potential \cite{deWit:1982ig}. The vacuum of interest preserves $\superN=2$ SUSY and the global symmetry $\grSU(3) \times \grU(1)$ (broken down from $\grSO(8)$) and corresponds to a scalar and a pseudo-scalar of $\superN=8$ gauge supergravity acquiring VEVs. As a consequence, this background does not preserve parity.

The 11d uplift of this $AdS_4$ vacuum was found more recently, in \cite{Corrado:2001nv}, and studied further in \cite{Ahn:2002eh}. The solution is not a simple Freund-Rubin direct product $AdS_4 \times X_7$ but instead the metric of $AdS_4$ is warped by a function $f(y)$ of the coordinates $y$ on the internal manifold $X_7$. $X_7$ itself is a `squashed and stretched' $S^7$ \cite{Corrado:2001nv}. As noted earlier, this background has an Englert type flux in the $S^7$ directions\cite{Englert:1982vs}, which is another way of seeing the breaking of parity.

To determine the SUGRA spectrum, one could in principle perform the $11 \to 4$ dimensional KK reduction on this warped, squashed and stretched space. By performing the KK reduction for modes of various $AdS_4$ spin, one can group the resulting particles into $\superN=2$ supermultiplets of definite energy. Such an analysis was performed for example in \cite{Gualtieri:1999tu,Fabbri:1999mk} for Freund-Rubin vacua with $X_7 = M^{111},Q^{111}$. We can avoid such an involved calculation for this warped spacetime since it is obtained at the end of a SUSY preserving RG flow from the $\superN=8$ theory. A similar analysis has been performed earlier in \cite{Nicolai:1985hs} for the $\grSU(3) \times \grU(1)_R$ case at hand and in \cite{Freedman:1999gp} for the analogous case in AdS$_5$/CFT$_4$ for a gauge theory with $\grSU(2) \times \grU(1)$ symmetry. However, here we go beyond gauged supergravity and study the rearrangement of the massive KK modes of the $\superN=8$ theory into $\superN=2$ supermultiplets. In doing so, we find that of the two alternative charge assignments, only one (referred to as ``Scenario I'' below) leads to agreement with the proposed gauge dual, while the other (``Scenario II'')
does not appear to be characteristic of a KK reduction. Hence while both assignments are consistent at the level of symmetry, only the former is likely to be reproduced through an explicit KK reduction from $11 \to 4$ dimensions.

\begin{table}[H]
\begin{center}
\begin{tabular}{c|c|c|c}
Spin & Field & $\grSO(8)$ irrep & $\grSO(8)$ Dynkin labels \\ \hline
$2$           & $e_\mu{}^a$    & $\rep{1}$    & $[0,0,0,0]$ \\
$\frac{3}{2}$ & $\psi_\mu{}^I$ & $\rep{8}_s$  & $[0,0,0,1]$ \\
$1$           & $A_\mu{}^{IJ}$ & $\rep{28}$   & $[0,1,0,0]$ \\
$\frac{1}{2}$ & $\chi^{IJK}$   & $\rep{56}_s$ & $[1,0,1,0]$ \\
$0^+$         & $S^{[IJKL]_+}$ & $\rep{35}_v$ & $[2,0,0,0]$ \\
$0^-$         & $P^{[IJKL]_-}$ & $\rep{35}_c$ & $[0,0,2,0]$
\end{tabular}
\caption{\textbf{\mathversion{bold} The massless $\superN = 8$ supermultiplet.} All degrees of freedom of 11d supergravity form one supermultiplet. When compactified on a round seven-sphere this supermultiplet splits into a series of $\grOsp(8|4)$ supermultiplets. This table lists the components of the lowest supermultiplet in this series.}
\label{tab:masslessN=8}
\end{center}
\end{table}

\begin{table}[H]
\begin{center}
\begin{tabular}{c|c|l}
Spin & Field & $\grSO(8)$ Dynkin labels \\ \hline
$2$           & $e_\mu{}^a$    & $[n,0,0,0]$ \\
$\frac{3}{2}$ & $\psi_\mu{}^I$ & $[n,0,0,1]+[n-1,0,1,0]$ \\
$1$           & $A_\mu{}^{IJ}$ & $[n,1,0,0]+[n-1,0,1,1]+[n-2,1,0,0]$ \\
$\frac{1}{2}$ & $\chi^{IJK}$   & $[n+1,0,1,0]+[n-1,1,1,0]+[n-2,1,0,1]+[n-2,0,0,1]$ \\
$0^+$         & $S^{[IJKL]_+}$ & $[n+2,0,0,0]+[n-2,2,0,0]+[n-2,0,0,0]$ \\
$0^-$         & $P^{[IJKL]_-}$ & $[n,0,2,0]+[n-2,0,0,2]$
\end{tabular}
\caption{\textbf{\mathversion{bold} The massive $\superN = 8$ supermultiplet at level $n$.} Representations with negative labels are absent. For $n=0$ the massless $\superN = 8$ supermultiplet from \protect\tabref{tab:masslessN=8} is recovered.}
\label{tab:massiveN=8}
\end{center}
\end{table}

\subsection{Spectrum on the stretched and squashed seven-sphere}

The spectrum of $\superN=8$ supermultiplets obtained by KK reduction on the round $\Sphere^7$ is well-known \cite{Duff:1986hr,Biran:1983iy}. All multiplets are shortened and have maximum spin $2$. The massless multiplet is shown in \tabref{tab:masslessN=8} while the $\grSO(8)_R$ representations that higher massive multiplets transform in is presented in \tabref{tab:massiveN=8}.

Now we would like to find the spectrum on the deformed $\Sphere^7$. We do this by exploiting the restrictions on the spectrum due to the symmetries of the background. 
The strategy for this derivation is summarized in the following diagram:
\be
\begin{CD}
\grOsp(8|4) @>\mbox{\hspace{5mm}\footnotesize stretching and\hspace{5mm}}>\mbox{\footnotesize squashing of $\Sphere^7$}> \grSU(3)\times\grOsp(2|4) \\
@V{\mbox{\footnotesize decompose $\superN=8$} \atop \mbox{\footnotesize supermultiplets}}VV   @AA{\mbox{\footnotesize assemble $\superN=2$} \atop \mbox{\footnotesize supermultiplets}}A \\
\grSO(8)_R\times\grSO(3,2) @>\mbox{\hspace{5mm}\footnotesize RG flow\hspace{5mm}}>> \grSU(3)\times\grU(1)_R\times\grSO(3,2)
\end{CD}
\ee
The $\grOsp(8|4)$ supermultiplets are decomposed into irreducible representations of the bosonic subgroup $\grSO(8)_R\times\grSO(3,2)$ as already given in \tabref{tab:masslessN=8} and \ref{tab:massiveN=8}. This set of representations is then further decomposed into irreducible representations of the bosonic symmetry group $\grSU(3)\times\grU(1)_R\times\grSO(3,2)$ of the IR theory. Finally, we reassemble these bosonic multiplets into supermultiplets of $\grOsp(2|4)$ with definite $\grSU(3)$ representations. This procedure is carried out for every level $n$ separately.

The described method is applicable because the RG flow preserves the $\grOsp(2|4)\subset\grOsp(8|4)$ supersymmetry. The only thing we do not know is how $\grOsp(2|4)$ is embedded into $\grOsp(8|4)$, or how $\grSU(3)\times\grU(1)_R\times\grSO(3,2)$ is embedded into $\grSO(8)_R\times\grSO(3,2)$. Therefore we will make a general ansatz for the latter embedding:
\be \label{eqn:embedding-ansatz}
  [a,b,c,d] \to [f,g]_{h} \; ,
\ee
where $f$, $g$, $h$ are linear functions of the $\grSO(8)_R$ Dynkin labels $a,b,c,d$. The functions $f$ and $g$ represent the $\grSU(3)$ Dynkin labels, and the function $h$ is the $\grU(1)_R$ charge. The $\grSO(3,2)$ labels are given by the spin $s$ and the energy $E$. While the spin is unaltered during the flow, the energy can in general not be determined by group theoretical arguments alone. We can only find the energy for short multiplets where it is fixed by the values of the other labels.

The functions in the ansatz \eqref{eqn:embedding-ansatz} are restricted in the following way. First of all there are only three choices of canonical embeddings of $\grSU(3)$ into $\grSO(8)$ which are given by $[f,g]=[a,b]$ or $[b,c]$ or $[b,d]$. All three choices lead to the same decomposition if the R-charge is ignored; the result for the massless level is printed in \tabref{tab:massless-decomp}. We can now fix the $\grU(1)_R$ charges as follows. Fields of different spin but same $\grSU(3)$ representation in the decomposition of the $\superN=8$ supermultiplet must all recombine into various $\superN=2$ supermultiplets which we list in \tabref{tab:massless-graviton} to \ref{tab:massive-hyper} in \appref{sec:supermultiplets}. This is only possible when the R-charges of the states that go into one $\superN=2$ supermultiplet are correlated in the way given in the tables.
\begin{table}
\begin{center}
\begin{tabular}{ccll}
Spin & $\grSO(8)$ &  &$\grSU(3)$ \\
$2$           & $\rep{1}$    & $\to$ &
 $\rep{1}$ \\
$\frac{3}{2}$ & $\rep{8}_s$  & $\to$ &
 $\rep{3} + \rep{\bar{3}} + 2\cdot\rep{1}$ \\
$1$           & $\rep{28}$   & $\to$ &
 $\rep{8} + 3\cdot\rep{3} + 3\cdot\rep{\bar{3}} + 2\cdot\rep{1}$ \\
$\frac{1}{2}$ & $\rep{56}_s$ & $\to$ &
 $2\cdot\rep{8} + \rep{6} + \rep{\bar{6}} + 4\cdot\rep{3} + 4\cdot\rep{\bar{3}} + 4\cdot\rep{1}$ \\
$0^+$         & $\rep{35}_v$ & $\to$ &
 $\rep{8} + \rep{6} + \rep{\bar{6}} + 2\cdot\rep{3} + 2\cdot\rep{\bar{3}} + 3\cdot\rep{1}$ \\
$0^-$         & $\rep{35}_c$ & $\to$ &
 $\rep{8} + \rep{6} + \rep{\bar{6}} + 2\cdot\rep{3} + 2\cdot\rep{\bar{3}} + 3\cdot\rep{1}$
\end{tabular}
\caption{\textbf{\mathversion{bold} Decomposition of the massless $\superN = 8$ supermultiplet under $\grSU(3)$.}}
\label{tab:massless-decomp}
\end{center}
\end{table}

For example, there are only three fields in \tabref{tab:massless-decomp} in the sextet $\rep{6}$ of $\grSU(3)$ -- a spin $1/2$ field and two scalars. (Recall that the deformation and the IR background are not parity invariant and hence the UV parity assignments should be ignored.) The only supermultiplet they can form is a hypermultiplet described by \tabref{tab:massive-hyper}. This requires an R-charge assignment of the form\footnote{We adopt the usual notation where the upper sign applies if $y_0>0$ and the lower one if $y_0<0$.} $y_0 \mp 1$ for the spin $1/2$ field when the scalars are assigned $y_0, y_0\mp 2$. When we repeat this multiplet-forming exercise for the other fields, this further constrains the embedding of $\grU(1)_R$ into $\grSO(8)_R$ until we are left with exactly two possibilities consistent with supersymmetry.

Doing this in a systematic way, we find that the two choices can be described in terms of the Dynkin labels of $\grSO(8)$ as,
 \be
  [a,b,c,d] \to
  \begin{cases}
  [a,b]_{ \lrbrk{\frac{a}{3} +\frac{2b}{3}  +d}\eps} & \mbox{Scenario I}  \; , \\
  [a,b]_{-\lrbrk{\frac{2a}{3}+\frac{4b}{3}+c+d}\eps} & \mbox{Scenario II} \; ,
  \end{cases}
\ee
where $\eps = \pm 1$, the two integers $[a,b]$ give the $\grSU(3)$ Dynkin labels and the subscript is the $\grU(1)_R$ charge. The choice of $\eps = \pm 1$ is simply a flip of the $\grU(1)_R$ definition. We note that the $\grSU(3)$ embedding $[b,c]$ and $[b,d]$ lead to no consistent regrouping into $\superN=2$ supermultiplets.

With the $\grSU(3) \times \grU(1)_R$ charges of Scenario I above, we proceed to group the fields into $\superN=2$ supermultiplets. The result for fields from the $\superN=8$ massless sector is in \tabref{tab:decomp-Betti}. We find from the table that the massless $\superN=8$ multiplet yields some familiar massless $\superN=2$ multiplets such as the massless graviton multiplet in a singlet under $\grSU(3)$ and a massless vector multiplet in the adjoint of $\grSU(3)$. The former is expected in any theory of SUGRA while the latter contains the massless bosons gauging the $\grSU(3)$ symmetry in the bulk. We also find several other massive multiplets that acquired mass in the breaking $\grSO(8) \rightarrow \grSU(3) \times \grU(1)_R$. When massless particles of spin $1$ or greater acquire a mass, they need to `eat' spin $1/2$ and spin $0$ particles to furnish the extra polarizations. Hence when we find massive gravitinos in a $\rep{3}$ of $\grSU(3)$ for example, we need to set aside some spin $1/2$ triplets from \tabref{tab:massless-decomp} to be eaten and not group them into other multiplets. These are listed in the last column of \tabref{tab:decomp-Betti}.

Scenario II produces a different set of $\grU(1)_R$ charges. The grouping of massless $\superN=8$ fields into $\superN=2$ multiplets is detailed in \tabref{tab:decomp-Warner}. The crucial differences between \tabref{tab:decomp-Warner} and \tabref{tab:decomp-Betti} are in the hyper- and long vector multiplets where a reassignment of R-charges leads to differing physical dimensions. For the hypermultiplet, Scenario II (\tabref{tab:decomp-Warner}) assigns the ground state a R-charge of $y_0 = - \frac{4}{3}$ and hence by \tabref{tab:massive-hyper}, a dimension of $E_0 = \abs{y_0} = \frac{4}{3}$. On the other hand, Scenario I (\tabref{tab:decomp-Betti}) results in the assignment $y_0 = \frac{2}{3}$ and $E_0 = \abs{y_0} = \frac{2}{3}$. In \appref{app:dressing} we further observe that the energies of the hypermultiplet in the two scenarios can be related to the same mass spectrum when different dressings are used for the two different scenarios. However, this relationship holds only at level $n=0$.

\begin{table}
\begin{center}
\begin{tabular}{|c|c|c|c|c|c|c|c|c|c|} \hline
Spin & $\grSO(8)$ & \multicolumn{8}{c|}{$\grSU(3)_{\grU(1)}$} \\ \hline
%
%
$2$ & $\rep{1}$ & $\rep{1}_{0}$ &&&&&&&   \\ \hline
%
%
$\frac{3}{2}$ & $\rep{8}_s$ & $\rep{1}_{+1}$ & & $\rep{3}_{\et}$ & $\rep{\bar{3}}_{-\et}$ &  &  && \\
&&$\rep{1}_{-1}$ & &&&&&&  \\  \hline
%
%
$1$ & $\rep{28}$ &  $\rep{1}_{0}$ & $\rep{8}_{0}$& $\rep{3}_{\eft}$ & $\rep{\bar{3}}_{-\eft}$ && & $\rep{1}_{0}$ & \\
&&&& $\rep{3}_{-\ett}$ & $\rep{\bar{3}}_{\ett}$ &&&& \\
&&&& $\rep{3}_{-\ett}$ & $\rep{\bar{3}}_{\ett}$  && && \\ \hline
%
%
$\frac{1}{2}$ & $\rep{56}_s$ & &$\rep{8}_{+1}$  & $\rep{3}_{\et}$ & $\rep{\bar{3}}_{-\et}$ & $\rep{6}_{-\et}$ & $\rep{\bar{6}}_{\et}$ & $\rep{1}_{-1}$ & $\rep{3}_{\et}$ \\  
&& &$\rep{8}_{-1}$ &$\rep{3}_{\et}$ & $\rep{\bar{3}}_{-\et}$ &&  & $\rep{1}_{+1}$ & $\rep{\bar{3}}_{-\et}$ \\ 
&&&   & $\rep{3}_{-\evt}$ & $\rep{\bar{3}}_{\evt}$ & & & $\rep{1}_{-1}$ & \\ 
&&&&&&& & $\rep{1}_{+1}$ & \\ \hline 
%
%
$0^+$ & $\rep{35}_v$ & & $\rep{8}_{0}$ &  $\rep{3}_{-\ett}$ & $\rep{\bar{3}}_{\ett}$  & $\rep{6}_{\ett}$ & $\rep{\bar{6}}_{-\ett}$ &  $\rep{1}_{2}$ & $\rep{3}_{\eft}$ \\
&&&&&&&& $\rep{1}_{0}$ & $\rep{\bar{3}}_{-\eft}$ \\
&&&&&&&&$\rep{1}_{-2}$ & \\ \hline
%
%
$0^-$ & $\rep{35}_c$ & &$\rep{8}_{0}$ & && $\rep{6}_{-\eft}$ & $\rep{\bar{6}}_{\eft}$ & $\rep{1}_{0}$ &  $\rep{3}_{-\ett}$ \\
&&&&&&&& $\rep{1}_{0}$ & $\rep{3}_{-\ett}$ \\
&&&&&&&&  & $\rep{\bar{3}}_{\ett}$ \\
&&&&&&&&& $\rep{\bar{3}}_{\ett}$ \\
&&&&&&&&& $\rep{1}_0$ \\
\hline
\multicolumn{2}{l|}{} &
\rotatebox{90}{\mbox{Massless graviton\;}} &
\rotatebox{90}{\mbox{Massless vector\;}} &
\rotatebox{90}{\mbox{Massive short gravitino\;}} &
\rotatebox{90}{\mbox{Massive short gravitino\;}} &
\rotatebox{90}{\mbox{Massive hyper\;}} &
\rotatebox{90}{\mbox{Massive hyper\;}} &
\rotatebox{90}{\mbox{Massive vector\;}} &
\rotatebox{90}{\mbox{eaten\;}} \\  \cline{3-10}
\end{tabular}
\caption{\textbf{\mathversion{bold} Decomposition of massless $\superN = 8$ supermultiplet: Scenario I.} $\eps$ can be set to $\pm1$.}
\label{tab:decomp-Betti}
\end{center}
\end{table}

\begin{table}
\begin{center}
\begin{tabular}{|c|c|c|c|c|c|c|c|c|c|} \hline
Spin & $\grSO(8)$ & \multicolumn{8}{c|}{$\grSU(3)_{\grU(1)}$} \\ \hline
%
%
$2$ & $\rep{1}$ & $\rep{1}_{0}$ &&&&&&&   \\ \hline
%
%
$\frac{3}{2}$ & $\rep{8}_s$ & $\rep{1}_{+1}$ & & $\rep{3}_{\et}$ & $\rep{\bar{3}}_{-\et}$ &  &  && \\
&&$\rep{1}_{-1}$ & &&&&&&  \\  \hline
%
%
$1$ & $\rep{28}$ &  $\rep{1}_{0}$ & $\rep{8}_{0}$& $\rep{3}_{\eft}$ & $\rep{\bar{3}}_{-\eft}$ && & $\rep{1}_{0}$ & \\
&&&& $\rep{3}_{-\ett}$ & $\rep{\bar{3}}_{\ett}$ &&&& \\
&&&& $\rep{3}_{-\ett}$ & $\rep{\bar{3}}_{\ett}$  && && \\ \hline
%
%
$\frac{1}{2}$ & $\rep{56}_s$ & &$\rep{8}_{+1}$  & $\rep{3}_{\et}$ & $\rep{\bar{3}}_{-\et}$ & $\rep{6}_{-\et}$ & $\rep{\bar{6}}_{\et}$ & $\rep{1}_{-1}$ & $\rep{3}_{\et}$ \\  
&& &$\rep{8}_{-1}$ &$\rep{3}_{\et}$ & $\rep{\bar{3}}_{-\et}$ &&  & $\rep{1}_{+1}$ & $\rep{\bar{3}}_{-\et}$ \\ 
&&&   & $\rep{3}_{-\evt}$ & $\rep{\bar{3}}_{\evt}$ & & & $\rep{1}_{-1}$ & \\ 
&&&&&&& & $\rep{1}_{+1}$ & \\ \hline 
%
%
$0^+$ & $\rep{35}_v$ & & $\rep{8}_{0}$ &  $\rep{3}_{-\ett}$ & $\rep{\bar{3}}_{\ett}$  & $\rep{6}_{-\eft}$ & $\rep{\bar{6}}_{\eft}$ &  $\rep{1}_{0}$ & $\rep{3}_{-\ett}$ \\
&&&&&&&& $\rep{1}_{0}$ & $\rep{\bar{3}}_{\ett}$ \\
&&&&&&&&& $\rep{1}_0$ \\ \hline
%
%
$0^-$ & $\rep{35}_c$ & &$\rep{8}_{0}$ & && $\rep{6}_{\ett}$ & $\rep{\bar{6}}_{-\ett}$ & $\rep{1}_{+2}$ &  $\rep{3}_{\eft}$ \\
&&&&&&&& $\rep{1}_{0}$ & $\rep{3}_{-\ett}$ \\
&&&&&&&& $\rep{1}_{-2}$ & $\rep{\bar{3}}_{-\eft}$ \\
&&&&&&&&& $\rep{\bar{3}}_{\ett}$ \\ \hline
\multicolumn{2}{l|}{} &
\rotatebox{90}{\mbox{Massless graviton\;}} &
\rotatebox{90}{\mbox{Massless vector\;}} &
\rotatebox{90}{\mbox{Massive short gravitino\;}} &
\rotatebox{90}{\mbox{Massive short gravitino\;}} &
\rotatebox{90}{\mbox{Massive hyper\;}} &
\rotatebox{90}{\mbox{Massive hyper\;}} &
\rotatebox{90}{\mbox{Massive vector\;}} &
\rotatebox{90}{\mbox{eaten\;}} \\  \cline{3-10}
\end{tabular}
\caption{\textbf{\mathversion{bold} Decomposition of massless $\superN = 8$ supermultiplet : Scenario II.}  $\eps$ can be set to $\pm1$.}
\label{tab:decomp-Warner}
\end{center}
\end{table}

The massive multiplets of $\superN=8$, listed in \tabref{tab:massiveN=8} for $n=1,2,3,\ldots$, are decomposed in a similar way for each of the two scenarios. We have delegated the details to the appendices in \tabref{tab:n0} through \ref{tab:n2-Warner}. We find several series of $\superN=2$ multiplets as we increase $n$ with different R-charges in the two different scenarios. To compare with the gauge theory, the short multiplets are the most interesting since their energy can be determined entirely from their R-charge. We collect the four distinct series of short multiplets that emerge from decomposing massive $\superN=8$ multiplets in \tabref{tab:short-series} for the two scenarios. The $\grSU(3)$ representations are given in terms of Dynkin labels, i.e.\ $[a,b]$ is the symmetric product of $a$ $\rep{3}$'s and $b$ $\rep{\bar{3}}$'s. The subscript again gives the $\grU(1)_R$ charge. When $n = 0$, these multiplets are also found in Tables \ref{tab:decomp-Warner},\ref{tab:decomp-Betti} discussed earlier.

We stress that the two scenarios arise as logical possibilities when one only works at the level of the symmetry breaking $\grSO(8) \to \grSU(3) \times \grU(1)_R$ and one does not perform an explicit KK reduction to find the mass spectrum. The two scenarios correspond to two different embeddings of $\grU(1)_R$ in $\grSO(8)$. From \tabref{tab:short-series}, one can easily verify that set of masses resulting from the two scenarios are distinct (though this is not true when $n=0$ as discussed in \appref{app:dressing}). For example, we note the unusual feature that in Scenario II, the short gravitons have the $n$-independent charge $[0,0]_0$. This leads to $n$-independent mass of $m^2 = 0$ for the graviton. It would seem very unlikely that such an infinite sequence of zero masses can be obtained from a KK reduction. In contrast, Scenario I has masses that increase with $n$ for all the short series. Thus Scenario II is unlikely to be obtained from an explicit KK reduction and we conjecture that it is Scenario I that will agree with such a direct computation. Hence we will primarily work with Scenario I in this paper and compare it with proposed dual gauge theory.

\begin{table}[H]
\begin{center}
\begin{tabular}{|c|l|l|} \cline{2-3}
\multicolumn{1}{l|}{} & Scenario I & Scenario II \\ \hline
Hyper     & $[n+2,0]_{\frac{n+2}{3}}$, $[0,n+2]_{-\frac{n+2}{3}}$ & $[n+2,0]_{-\frac{2n+4}{3}}$, $[0,n+2]_{\frac{2n+4}{3}}$ \\
Vector    & $[n+1,1]_{\frac{n}{3}}$,   $[1,n+1]_{-\frac{n}{3}}$ & $[n+1,1]_{-\frac{2n}{3}}$,   $[1,n+1]_{\frac{2n}{3}}$ \\
Gravitino & $[n+1,0]_{\frac{n+1}{3}}$, $[0,n+1]_{-\frac{n+1}{3}}$ & $[n+1,0]_{-\frac{2n-1}{3}}$, $[0,n+1]_{\frac{2n-1}{3}}$ \\
Graviton  & $[0,0]_{n}$, $[0,0]_{-n}$ & $[0,0]_{0}$, $[0,0]_{0}$ \\ \hline
\end{tabular}
\caption{\textbf{\mathversion{bold} Series of short multiplets in the two scenarios.} These are four series of short multiplets labeled by $n=0,1,2,\ldots$.
When $n=0$, there is only one $[1,1]_0$ vector and one $[0,0]_0$ graviton, both of which are massless.}
\label{tab:short-series}
\end{center}
\end{table}

\section{Gauge theory side}
\label{sec:GaugeTheory}

In this section we discuss the conjectured gauge theory dual to the supergravity background described above, i.e. the warped product of $\AdS_4$ and a squashed and stretched $\Sphere^7$. We provide evidence that this gauge theory is the IR limit of the ABJM theory \cite{Aharony:2008ug} with a superpotential mass term for one of the superfields, as conjectured in \cite{Benna:2008zy} (see also \cite{Ahn:2008ya}).

\subsection{Review of ABJM theory}

We begin with a brief recap of ABJM theory \cite{Aharony:2008ug} following the notation in \cite{Benna:2008zy}. The $\grU(N)\times\grU(N)$ gauge superfields are $\VV^a{}_b$ and $\hat{\VV}^\ha{}_\hb$. The matter superfields $(\ZZ^A)^a{}_\ha$ and $(\WW_A)^\ha{}_a$ transform under gauge transformations in the representation $(\rep{N},\rep{\bar{N}})$ and $(\rep{\bar{N}},\rep{N})$, respectively. They also transform under two different global $\grSU(2)$'s in the $\rep{2}$ and $\rep{\bar{2}}$ indicated by the indices $A=1,2$. The action is given by standard Chern-Simons terms with level $k$ for $\VV$ and level $-k$ for $\hat{\VV}$. The matter action is given by the standard kinetic terms for $\ZZ$ and $\WW$ minimally coupled to the gauge fields. Finally, the theory includes the $\grSU(2)\times \grSU(2)$ invariant superpotential \cite{Klebanov:1998hh}
\be \label{KWsup}
  \superW = \frac{1}{4} \levi_{AC}\levi^{BD} \tr \ZZ^A \WW_B \ZZ^C \WW_D \; .
\ee
This gauge theory was conjectured to be the CFT dual of M-theory on $\AdS_4\times(\Sphere^7/\Integers_k)$ supported by $N$ units of 4-form flux \cite{Aharony:2008ug}.

Special attention needs to be paid to the $\grU(1)\times \grU(1)$ part of the gauge group. All matter fields are neutral under one linear combination of the $\grU(1)$'s, $c_\mu$, which therefore corresponds to the center of mass degree of freedom of the stack of M2-branes. The flux for this non-interacting $\grU(1)$ is quantized, and it may be dualized into a periodic scalar. As a result the other linear combination, the `baryonic' $\grU(1)$ gauge field $b_\nu$, which enters the Chern-Simons action as
\be \label{abelcs}
\frac{k}{2\pi} \levi^{\mu \nu \lambda} b_\mu \partial_\nu c_\lambda \; ,
\ee
gets broken to $\Integers_k$ \cite{Aharony:2008ug}
through a mechanism demonstrated in \cite{Lambert:2008et,Distler:2008mk}.
The generator of this group acts on the superfields as
\be
\ZZ^A \rightarrow e^{2\pi i/k} \ZZ^A\ , \qquad \WW_B \rightarrow e^{-2\pi i/k} \WW_B \; .
\ee
This argument loosely suggests that for $k=1$ the $\grU(N)\times \grU(N)$ gauge theory is simply
equivalent to the $\grSU(N)\times \grSU(N)$. However, this is not quite correct since the moduli spaces of the two theories are different \cite{Lambert:2008et,Distler:2008mk,Aharony:2008ug}.\footnote{For $k=2$ and $N=2$ the moduli spaces of the two gauge theories do coincide \cite{Lambert:2008et,Distler:2008mk}, so in this case they may be equivalent.} The ABJM theory with $k>2$ has been demonstrated to possess the $\superN=6$ superconformal invariance, in agreement with that of its proposed M-theory dual. For $k=1,2$ the superconformal symmetry of the ABJM theory is expected to enhance to $\superN=8$ but this is yet to be demonstrated explicitly. An important manifestation of this enhancement is that, when $\superN=2$ superspace is used, then the theory possesses $\grU(1)_R\times \grSU(4)$ global symmetry. Although the non-R $\grSU(4)$ flavor symmetry is difficult to establish in general, in the next section we discuss how it may appear using some specific examples.

\subsection{Towards establishing the $\grU(1)_R\times \grSU(4)$ invariance}

In \cite{Benna:2008zy} the BLG theory was reformulated using $\superN=2$ superspace where its has manifest $\grU(1)_R\times \grSU(4)$ global symmetry. This theory is exactly equivalent to the $\grSU(2)\times \grSU(2)$ version of the ABJM theory \cite{Aharony:2008ug}.
The four complex bi-fundamental superfields $\ZZ^A$ of the BLG theory, which transform in the fundamental of the global $\grSU(4)$, are related to the fields entering (\ref{KWsup}) through
\be \label{eqn:BLGZ3Z4}
  \ZZ^3 = \WW_1^\ddagger,\qquad \ZZ^4 = \WW_2^\ddagger \; .
\ee
This uses an operation special to the $\grSU(2)\times \grSU(2)$ gauge theory
\be \label{eqn:ddagger}
\WW^\ddagger := - \levi \WW^T \levi
\qquad
\mbox{with $\levi = \matr{cc}{0&1\\-1&0}$}
\ee
because it relies on the invariant tensor $\levi_{ab}$. After this transformation the superpotential (\ref{KWsup}), which is manifestly only $\grSU(2)^2$ invariant, acquires the $\grSU(4)$ invariant form \cite{Benna:2008zy}
\be
\superW= \frac{1}{4!} \levi_{ABCD} {\rm tr} \ZZ^A \ZZ^{\ddagger B} \ZZ^C  \ZZ^{\ddagger D} = - \frac{1}{8 \cdot 4!} \levi_{ABCD} \levi^{abcd} \ZZ^A_c \ZZ^{B}_b \ZZ^C_c  \ZZ^{D}_d
\ee
where the relation to $\grSO(4)$ notation $\ZZ^A_a$ is explained in \cite{Benna:2008zy}.

Below we will suggest how the operation (\ref{eqn:BLGZ3Z4}) may be generalized to $\grU(N)\times \grU(N)$ gauge theories. To accomplish this one likely has to invoke monopole operators, often called 't Hooft operators because of his pioneering work \cite{Hooft:1977hy}. Such operators naturally carry the magnetic charge determined by the flux they insert at a point. In a Chern-Simons theory, they also carry an electric charge (or gauge representation) proportional to the Chern-Simons level $k$. We assemble some useful facts about these operators in \appref{app:monopole}. For a recent explicit study of monopole operators in the ABJM theory, see \cite{Berenstein:2008dc}.

When $k=1$, the simplest monopole operators are $(e^\tau)^{a}_{\ha}$, which transforms in the representation ($\rep{N},\rep{\bar{N}}$), and its conjugate $(e^{-\tau})^{\ha}_{a}$. They are obtained for the choices of flux described in \appref{app:monopole}. We can also construct the ``double'' monopole operators, $(e^{2\tau})^{ab}_{\ha\hb}$ and $(e^{-2\tau})^{\ha\hb}_{ab}$. They can be either symmetric or anti-symmetric under separate interchanges of upper or lower indices, but both choices have the symmetry under the interchange of both:
\be
  (e^{2\tau})^{ab}_{\ha\hb} = (e^{2\tau})^{ba}_{\hb\ha}
  \comma
  (e^{-2\tau})^{\ha\hb}_{ab} = (e^{-2\tau})^{\hb\ha}_{ba}
  \; .
\ee
These operators transform under $\grU(N)\times\grU(N)$ as indicated by their indices. In particular, they are charged under the baryonic $\grU(1)$ gauge group, which is the interacting part of the $\grU(1)\times \grU(1)$. In our notation, $e^{n \tau}$ has charge $n$ under this baryonic $\grU(1)$.

When $k=2$, no choice of flux can give an operator of the form $(e^\tau)^{a}_{\ha}$ and the smallest operators one can form are $(e^{2 \tau})^{ab}_{\ha \hb}$ and its conjugate, as discussed in the appendix.

Let us use the monopole operators to establish $\grU(1)_R \times \grSU(4)$ symmetry of the $\grU(2)\times \grU(2)$ ABJM theory, which has some subtle differences from the BLG theory. Inspired by (\ref{eqn:ddagger}), we propose to use the monopole operators in the ABJM gauge theory that are \emph{ anti-symmetric} in each set of indices,
\be
 (e^{2\tau})^{ab}_{\ha\hb} = - (e^{2\tau})^{ba}_{\ha\hb}
	\comma
 (e^{2\tau})^{ab}_{\ha\hb} = - (e^{2\tau})^{ab}_{\hb\ha}
\; .
\ee
Thinking of the $N=2$ ABJM gauge theory as $\grSU(2)\times\grSU(2)\times \grU(1)\times \grU(1)$, we can use the  $\grSU(2)$ invariant tensors to write this as,
\be \label{Hooftexpl}
  (e^{2\tau})^{ab}_{\ha\hb} = T^2 \levi^{ab} \levi_{\ha \hb} \ , \quad (e^{-2\tau})^{\ha\hb}_{ab} = T^{-2} \levi_{ab} \levi^{\ha \hb} \ ,
\ee
where $T^2$ is a monopole operator that creates two (one) units of magnetic flux when $k=1$ ($k=2$) for the decoupled $\grU(1)$ field $c_\mu$ in the $\grU(1)\times\grU(1)$ Chern-Simons gauge theory \eqref{abelcs} coupled to the charged matter. Due to the coupling \eqref{abelcs}, $T^2$ is doubly charged under the baryonic $\grU(1)$ in both cases ($k=1,2$).

Using the expressions \eqref{Hooftexpl} valid when $N=2$, the following invertibility identity can be verified,
\be
  (e^{2\tau})^{ab}_{\ha\hb} (e^{-2\tau})^{\hb\hc}_{bc} = \delta^a_c \delta^\hc_\ha \; ,
\ee
where we have assumed that these monopole operators do not contribute to 
the scaling dimensions of gauge invariant operators. This is a non-trivial assumption, since in some theories where the monopole operators were constructed explicitly, their scaling dimensions are non-vanishing \cite{Borokhov:2002cg,Borokhov:2003yu}. The assumption that their scaling dimensions vanish in the ABJM theory was central in forming operators with the right dimension and R-charge for AdS/CFT duality \cite{Aharony:2008ug}, and that will be the case here as well. However a definitive proof of this has been lacking.

To search for a global symmetry enhancement in the superpotential \eqref{KWsup}, let us introduce a  multiplet of superfields in the
fundamental of $\grSU(4)$
\be
  \ZZ^A = (\ZZ^1, \ZZ^2, \WW_1 e^{2\tau}, \WW_2 e^{2\tau}) \; , \qquad A=1,2,3,4,
\ee
where the explicit index structure is
\be \label{eqn:largeNZ3Z4}
 (\ZZ^3)^a{}_\ha = (\WW_1)^{\hb}{}_b (e^{2\tau})^{ab}_{\ha \hb}
 \comma
 (\ZZ^4)^a{}_\ha = (\WW_2)^{\hb}{}_b (e^{2\tau})^{ab}_{\ha \hb}.
\ee
We note that the fields $\ZZ^A, A=1,\ldots 4$ have the same baryonic charge, even though $\WW_{1,2}$ have the opposite charge. With this definition the superpotential can be written as
\be\label{superpotential}
  \superW = \Half (\ZZ^1)^a{}_\ha (\ZZ^2)^b{}_\hb (\ZZ^3)^c{}_\hc (\ZZ^4)^d{}_\hd
            \Bigsbrk{ (e^{-2\tau})^{\ha\hc}_{bc} (e^{-2\tau})^{\hb\hd}_{ad}
                    - (e^{-2\tau})^{\ha\hd}_{bd} (e^{-2\tau})^{\hb\hc}_{ac} }
  \; .
\ee
In the $\grU(2)\times \grU(2)$ ABJM theory, using the expressions \eqref{Hooftexpl}, we find that the superpotential \eqref{superpotential} has a close relation to that of the BLG theory, but also contains the abelian monopole operators needed for its $\grU(1)_b$ gauge invariance:
\be
\superW= \frac{1}{4!} T^{-4} \levi_{ABCD} {\rm tr} \ZZ^A \ZZ^{\ddagger B} \ZZ^C  \ZZ^{\ddagger D} \ .
\ee
It would be very interesting to extend the validity of the above arguments and expressions to $N > 2$, and to establish the $\grSU(4)$ invariance of the superpotential in the $\grU(N)\times \grU(N)$ ABJM theory. This would provide a clear argument in favor of its $\superN=8$ supersymmetry.

\subsection{Quadratic Deformations of the Superpotential}

While for any $k$ we can add quadratic operators of the form $ \tr \ZZ^A \WW_B $,
for $k=1$ and $k=2$ we can also deform the ABJM theory by a relevant operator which is quadratic in
{\it just one} of the chiral superfields.
 To write these operators explicitly we need the monopole operators:
\be
\Delta \superW = m (\ZZ^4)^{a}{}_\ha (\ZZ^4)^{b}{}_\hb (e^{-2\tau})^{\ha\hb}_{a b}
\ee
This relevant operator creates RG flow. To find the effective superpotential of the infrared theory, we integrate out the massive field $\ZZ^4$ in the IR, leaving a sextic potential for the remaining fields. It is natural to conjecture \cite{Benna:2008zy} that this IR fixed point is dual to the warped $AdS_4$ background of M-theory containing a $\grU(1)_R\times \grSU(3)$ symmetric `squashed and stretched' 7-sphere, whose original gauged supergravity formulation was found in \cite{Warner:1983vz}.
In order to achieve the $\grU(1)_R$ symmetry, the total R-charge of the superpotential should equal $2$. In \cite{Aharony:2008ug} it was assumed that all the necessary monopole operators have vanishing R-charge and dimension. We will assume the same here without a more detailed study involving matter fields to justify this. Then we can assign the following dimensions and R-charges:
\be
\Delta(\ZZ^A) = R(\ZZ^A) = \frac{1}{3} \qquad \mbox{for $A=1,2,3$}
\comma
\Delta(\ZZ^4) = R(\ZZ^4) = 1 \; .
\ee
It is interesting that the $\grU(1)$ symmetry with these charges holds not just in the IR, but along the entire RG flow.
The M-theory dual of this RG flow was found in \cite{Warner:1983vz,Ahn:2002eh}. Remarkably, it possesses \cite{Johnson:2001ze} a $\grU(1)$ symmetry with the same charges as in the field theory.\footnote{We thank Juan Maldacena for an enlightening discussion on this issue.} This can be demonstrated by identifying the $\grU(1)$ symmetry of the 3-form potential (see eq. (121), (122) of \cite{Johnson:2001ze}) and showing that three of the complex coordinates of the 7-sphere transform with charge $1/3$, and the fourth one with charge $1$. This provides an immediate check of the gauge/gravity duality along the entire RG flow.

For general $N$ explicit demonstration of the $\grSU(3)$ global symmetry of the superpotential remains a challenge, just like the $\grSU(4)$ global symmetry of the ABJM superpotential \eqref{superpotential}. Fortunately, this symmetry is explicit for $\grU(2)\times \grU(2)$ ABJM theory, if we use our assumption \eqref{Hooftexpl} about the monopole operators. Then the quadratic superpotential deformation assumes the form
\be \label{eqn:deform-N=2}
  \Delta \superW = m T^{-2} \tr \ZZ^4 \ZZ^{4\ddagger} \; .
\ee
which is closely related to the deformation of the BLG theory proposed in \cite{Benna:2008zy}. Adding such a mass term and integrating out $\ZZ^4$, we find
\be
\ZZ^4 = - \frac{T^{-2}}{12m} \, \levi_{ABC} \, \ZZ^A \ZZ^{\ddagger B} \ZZ^C
\ee
and hence the new superpotential,
\be
\superW_{\mathrm{eff}} = \frac{T^{-6}}{144m} \,
    \levi_{ABC} \levi_{DEF} \,
    \tr \ZZ^A \ZZ^{\ddagger B} \ZZ^C \ZZ^{\ddagger D} \ZZ^E \ZZ^{\ddagger F} \; .
\ee

\bigskip
We conclude this section by making the breaking of parity invariance due to the deformation \eqref{eqn:deform-N=2} more apparent\footnote{We use the notation and conventions of \cite{Benna:2008zy}.}. The parity operation in the gauge theory sends $(x^0,x^1,x^2) \rightarrow (x^0,-x^1,x^2)$ \cite{Bandres:2008vf}. The fermionic coordinates transform as $\theta_\alpha = - \gamma^1_{\alpha\beta} \theta^\beta$. These maps are accompanied by a transformation of the fields. In the $N=2$ theory the superfield transforms as $\ZZ^A \rightarrow \ZZ^{\ddagger A}$ ($A=1,\ldots,4$), and the component fields as $Z^A \rightarrow Z^{\ddagger A}$, $\zeta^A_\alpha \rightarrow \gamma^1_{\alpha\beta} \zeta^{\ddagger A \beta}$, and $F^A \rightarrow -F^{\ddagger A}$ ($A=1,\ldots,4$). Now, consider the deformation \eqref{eqn:deform-N=2} integrated over superspace
\be
 \Delta \Lagr_{\mathrm{pot}} \eq m T^{-2} \int\!d^2\theta\: \tr \ZZ^4 \ZZ^{\ddagger 4} \nn \\
 \eq - m T^{-2} \tr \zeta^4 \zeta^{\ddagger 4} + 2m T^{-2} \tr F^4 Z^{\ddagger 4} \nn \\
 \eq - m T^{-2} \tr \zeta^4 \zeta^{\ddagger 4} + T^{-4} \frac{mL}{3} \levi^{ABC} \tr \bar{Z}_A \bar{Z}_B^\ddagger \bar{Z}_C Z^{\ddagger 4} \; ,
 \label{eqn:mass-deform-Lagr}
\ee
where in the last line we replaced the auxiliary field $F$ using its equation of motion. Any of these expressions makes it explicit that $\Delta \Lagr_{\mathrm{pot}}$ is parity odd and hence breaks the parity invariance of the original theory.

\section{Matching of short multiplets}
\label{sec:Matching}

Having described the field content of the IR fixed point, we can proceed to match gauge theory operators with the gravity multiplets found earlier. For every supermultiplet there is a superfield of the gauge theory. Long supermultiplets correspond to unconstrained superfields, short supermultiplets to constrained ones. We will focus on the four series of short multiplets (cf. \tabref{tab:short-series}) and show that with our assignment there are four corresponding series of gauge theory operators. For the duality to hold it is essential to assign the charges of the IR gravity states according to Scenario I.

To facilitate the comparison of the components of the gravity supermultiplets and the components of the gauge theory superfields, we summarize the charges of the component fields in \tabref{tab:charges}.
\begin{table}[H]
\begin{center}
\begin{tabular}{c|c|c|c|c|c|c|c|c|c|c|c|c}
           & $Z^A$          & $\zeta^A$      & $Z^\dagger_A$     & $\zeta^\dagger_A$ &
             $Z^4$          & $\zeta^4$      & $Z^\dagger_4$     & $\zeta^\dagger_4$ &
             $x$            & $\theta$       & $\bar{\theta}$  &                 \\ \hline
$\grSU(3)$ & $\rep{3}$      & $\rep{3}$      & $\bar{\rep{3}}$ & $\bar{\rep{3}}$ &
             $\rep{1}$      & $\rep{1}$      & $\rep{1}$       & $\rep{1}$       &
             $\rep{1}$      & $\rep{1}$      & $\rep{1}$       &                 \\
Dimension  & $\frac{1}{3}$  & $\frac{5}{6}$  & $\frac{1}{3}$   & $\frac{5}{6}$   &
             $1$            & $\frac{3}{2}$  & $1$             & $\frac{3}{2}$   &
             $-1$           & $-\frac{1}{2}$ & $-\frac{1}{2}$  &                 \\
R-charge   & $+\frac{1}{3}$ & $-\frac{2}{3}$ & $-\frac{1}{3}$  & $+\frac{2}{3}$  &
             $+1$           & $0$            & $-1$            & $0$             &
             $0$            & $+1$           & $-1$            &
\end{tabular}
\caption{\textbf{Dimensions and R-charges of building blocks.} The components of the superfields are $\ZZ = Z + \sqrt{2} \theta^\alpha \zeta_\alpha + \mbox{aux.}$ and $\bar{\ZZ} = Z^\dagger - \sqrt{2} \bar{\theta}^\alpha \zeta^\dagger_\alpha + \mbox{aux.}$}
\label{tab:charges}
\end{center}
\end{table}

\subsubsection*{Hypermultiplets}

In \secref{sec:sugra} we found that, in Scenario I, the hypermultiplets come in the $\grSU(3)$ representations $[n+2,0]$ where $n=0,1,2,\ldots$ (see left column of \tabref{tab:short-series}). They have R-charge $y_0 = \frac{n+2}{3}$ and dimension\footnote{ $y_0$ and $\Delta_0$ in this section must be compared to $y_0$ and $E_0$ in \tabref{tab:massless-graviton} to \ref{tab:massive-hyper}. Note that $\Delta_0=E_0$ refers to the dimension of the ground state in a multiplet and the dimensions of the other components are related as shown in those tables.} $\Delta_0 = \abs{y_0} = \frac{n+2}{3}$, both of which suggestively increase in steps of $1/3$, the R-charge and dimension of the superfields $\ZZ^A, A=1,2,3$. Hence we write down a series of corresponding operators,
\be \label{hyperop}
  H^{(n) A_1 \ldots A_{n+2}} \sim \ZZ^{(A_1} \ZZ^{A_2} \cdots \ZZ^{A_{n+2})} \; ,
\ee
ignoring their gauge indices for the moment. We have symmetrized the $\grSU(3)$ indices $A_i$ to obtain the $[n+2,0]$ representation. These operators are chiral, $\bar{D}_\alpha H^{(n)} = 0$, which implies that they have the structure of the $\superN=2$ hypermultiplet as given in \tabref{tab:massive-hyper}. To see this explicitly in this simple example, we write out the components of this superfield:
\be
  H^{(n)} &\sim& Z^{(A_1} \cdots Z^{A_{n+2})} \nl
          + n \, \sqrt{2} \, \theta^\alpha \, \zeta^{(A_1}_\alpha Z^{A_2} \cdots Z^{A_{n+2})} \nl
          - \half n(n-1) \, \theta^2 \, \zeta^{\alpha(A_1} \zeta^{A_2}_\alpha Z^{A_3} \cdots Z^{A_{n+2})}
          \; .
\ee
Using the charges from \tabref{tab:charges}, it is simple to verify that the dimensions and R-charges, as well as the spins, of the components match.

To render the schematic operator expression \eqref{hyperop} gauge invariant, we need to make use of monopole operators. For even $n$, the natural expression is
\be
  H^{(n) A_1 \ldots A_{n +2}} = \tr \ZZ^{(A_1} \ZZ^{A_2} e^{-2\tau} \, \ZZ^{A_3} \ZZ^{A_4} e^{-2\tau}  \cdots \ZZ^{A_{n+1}} \ZZ^{ A_{n+2})} e^{-2\tau} \; ,
\ee
where the operator $e^{-2\tau}$ is contracted with the preceding field as $(\ZZ e^{-2\tau})^\ha{}_a = \ZZ^b{}_\hb (e^{-2\tau})^{\ha\hb}_{ab}$. For $N=2$, where the form of the monopole operators simplifies, these operators become
\be
  H^{(n) A_1 \ldots A_{n+2}} = T^{-n-2} \tr \ZZ^{(A_1} \ZZ^{\ddagger A_2} \, \ZZ^{A_3} \ZZ^{\ddagger A_4} \cdots \ZZ^{A_{n+1}} \ZZ^{\ddagger A_{n+2})} \; .
\ee
They are generalizations of the $n=0$ quadratic operator studied in \cite{Benna:2008zy}. In order to write down the operators for odd $n$, present for $k=1$, we need to insert one monopole operator $(e^{-\tau})^{\ha}_{a}$:
\be
  H^{(n) A_1 \ldots A_{n +2}} = \tr \ZZ^{(A_1} \ZZ^{A_2} e^{-2\tau} \cdots \ZZ^{A_{n}} \ZZ^{A_{n+1}} e^{-2\tau} \ZZ^{A_{n+2})} e^{-\tau}\; .
\ee
For $k=2$ the operator $e^{-\tau}$ is not available, and we can construct only the even operators. This is consistent with the supergravity side: when $n$ is odd and $k=2$, the $Z_2$ orbifold action projects out the corresponding SUGRA mode.

\subsubsection*{Short graviton multiplets}

From \tabref{tab:short-series}, we see that the short graviton multiplets are always $\grSU(3)$ singlets. In Scenario I they possess R-charges $y_0 = n$ and dimensions $\Delta_0 = \abs{y_0} + 2 = n + 2$ for $n=0,1,2,\ldots$. When $n=0$, this is actually the familiar massless graviton in $AdS$ and hence corresponds to the energy momentum tensor in the CFT. The other two massless components in this supermultiplet are the gravitino which is the SUSY generator and a massless vector boson which corresponds to the $\grU(1)_R$ symmetry of the dual CFT.

The gauge theory operator dual to the massless graviton multiplet is given by the stress-energy superfield
\be
  \mathcal{T}^{(0)}_{\alpha\beta} = \tr \bar{D}_{(\alpha} \bar{\ZZ}_A D_{\beta)} \ZZ^A + i \tr \bar{\ZZ}_A \!\stackrel{\leftrightarrow}{\partial}_{\alpha\beta}\! \ZZ^A \; ,
\ee
which satisfies the corresponding constraint $D^\alpha \mathcal{T}^{(0)}_{\alpha\beta} = \bar{D}^\alpha \mathcal{T}^{(0)}_{\alpha\beta} = 0$ and has protected classical dimension. For example the spin-two component has exact dimension 3 and the ground state component has dimension $\Delta_0=2$. For higher $n$ we expect the series to continue schematically as
\be \label{eqn:SGRAV-operator-schematic}
  \mathcal{T}^{(n)}_{\alpha\beta} \sim \mathcal{T}^{(0)}_{\alpha\beta} (\levi_{ABC} \ZZ^A \ZZ^B \ZZ^C)^n \qquad \mbox{for $n=1,2,3,\ldots$} \; ,
\ee
where we again understand none of the gauge indices to be contracted yet. The anti-symmetric combination of three $\ZZ$s may be thought of as the field $\ZZ^4$ which was integrated out. For $n\ge1$ these superfields satisfy only $\bar{D}^\alpha \mathcal{T}^{(n)}_{\alpha\beta} = 0$. Such a series has R-charge and dimension increasing in steps of $1$ and in complete agreement with Scenario I in \tabref{tab:short-series}.

The fields \eqref{eqn:SGRAV-operator-schematic} are again made gauge invariant by means of appropriate mo\-no\-pole operators. For even $n$ we insert a total of $3\frac{n}{2}$ monopole operators with two units of flux, $e^{-2\tau}$, and contract them with every other field as we described for the hypermultiplet. To find the superfield corresponding to the short graviton multiplet, one also needs to sum over all permutations of the fields. A typical term in such a sum is
\be
 \tr \mathcal{T}^{(0)}_{\alpha\beta} \lrsbrk{ \lrbrk{\levi_{ABC} \ZZ^{A}e^{-2 \tau} \ZZ^{B} \ZZ^{C}e^{-2 \tau} }  \lrbrk{ \levi_{DEF} \ZZ^{D} \ZZ^{E} e^{-2 \tau} \ZZ^{F} } \cdots } \; .
\ee
For odd $n$ we need to insert another monopole operator with one unit of flux, $e^{-\tau}$. If $k=2$ we do not have such a monopole at our disposal and hence there are no gauge theory operators for odd $n$. This mirrors the fact that such modes are projected out by the orbifolding action on the gravity side, just as we saw for the hypermultiplets.

The dimensions and R-charge in Scenario II appear difficult to interpret in a CFT. The corresponding short graviton series has a fixed R-charge of $0$ and dimension of $2$ for all $n$. As remarked earlier, this does not seem characteristic of a KK reduction.

\subsubsection*{Short gravitino multiplets}

The short gravitino multiplets come in the $\grSU(3)$ representations $[n+1,0]$ with R-charges $y_0 = \frac{n+1}{3}$ and dimensions $\Delta_0 = \abs{y_0} + \frac{3}{2} = \frac{2n+11}{6}$ for $n=0,1,2,\ldots$. Note that this is a massive multiplet even for $n=0$. The existence of a massless gravitino multiplet would indicate enhancement of SUSY beyond $\superN = 2$. Based on this data, we can write down the following candidate superfield,
\be \label{eqn:SGINO-operator-schematic}
  \Lambda^{(n) A_1 \ldots A_{n+1}}_\alpha \sim \levi_{ABC} \ZZ^A \ZZ^B \ZZ^C D_\alpha \ZZ^{(A_1} \ZZ^{A_2} \cdots \ZZ^{A_{n+1})}\; ,
\ee
where the derivative acts only onto the $\ZZ$ next to it. These fields are a fermionic superfields and satisfy $\bar{D}^\alpha \Lambda_\alpha = 0$. We can verify that \eqref{eqn:SGINO-operator-schematic} is the correct dual operator by checking the explicit components of this superfield against the known SUGRA multiplet. We show this for $n=0$. Let us restrict ourselves to $N=2$ where we can use the $\grSO(4)$ notation $\ZZ^A_a$
 that enables us to write the operator in the following gauge invariant way
\be
  \Lambda^{(0)A_1}_\alpha \sim \levi_{ABC} \levi^{abcd} \ZZ^A_a  \ZZ^B_b \ZZ^C_c D_\alpha \ZZ^{A_1}_{d} \;  \; .
\ee
The component expansion of this superfield is (up to total derivatives)
\begin{equation}
\begin{split}
  \Lambda^{(0)A_1}_\alpha \sim \levi_{ABC} \levi^{abcd} \Bigsbrk{
  &- \sqrt{2}i \, (\theta\gamma^\mu\bar{\theta}) \; \bigbrk{ Z Z Z \partial_\mu \zeta_\alpha
                         + \levi_{\mu\nu\rho} Z Z Z (\gamma^\nu\partial^\rho\zeta)_\alpha  \\
  &\hspace{26mm}         + 3 \, \zeta_\alpha Z Z \partial_\mu Z
                         - 3 \,  \levi_{\mu\nu\rho} (\gamma^\nu\zeta)_\alpha Z Z  \partial^\rho Z } \\
  &+ 2i \, (\gamma^\mu\bar{\theta})_\alpha \; Z Z Z \partial_\mu Z  \\
  &- 6i \, \theta^2 (\gamma^\mu\bar{\theta})_\alpha \; \bigbrk{ \zeta \zeta Z \partial_\mu Z
                                                           + Z Z \zeta \partial_\mu \zeta } \\
  &- 3 \, (\gamma^\mu\theta)_\alpha \; Z Z \zeta \gamma_\mu \zeta \\
  &- \sqrt{2}i \, \theta \bar{\theta} \; \bigbrk{ Z Z Z (\slashed{\partial} \zeta)_\alpha
                                            + 3 \,(\gamma^\mu \zeta)_\alpha Z Z  \partial_\mu Z } \\
  &+ \sqrt{2} \; Z Z Z  \zeta_\alpha \\
  &- 3 \, \sqrt{2} \theta^2 \; Z \zeta \zeta \zeta_\alpha \\
  &- 3 \, \theta_\alpha \; Z Z \zeta \zeta
  } \; .
\end{split}
\end{equation}
To simplify the notation, we have omitted the $\grSU(3)$ indices $ABCA_1$ and the $\grSO(4)$ gauge indices $abcd$ from the fields on the right hand side. The dimensions, R-charge and spin of each component presented on distinct lines above match up with the components of the supermultiplet in \tabref{tab:short-gravitino}.

The monopole operators required to make these operators gauge invariant for general $n$ are similar to those used for the hypermultiplets with $e^{-2 \tau}$ inserted on every other $\ZZ$ and summing over all permutations. A typical term in such a sum (when $n$ is even) is,
\be
\lrbrk{ \levi_{ABC} \ZZ^A e^{-2\tau} \ZZ^B  \ZZ^C e^{-2\tau} } D_\alpha \ZZ^{(A_1} \ZZ^{A_2} e^{-2\tau} \cdots \ZZ^{A_{n-1}} \ZZ^{A_n} e^{-2\tau} \ZZ^{A_{n+1})}
\ee

If $n$ is odd, we need an extra $e^{-\tau}$ monopole operator which is allowed only when $k=1$. This agrees with the fact that the corresponding SUGRA modes are projected out by the $k=2$ orbifold.

\subsubsection*{Short vector multiplets}

The short vector multiplets come in the $\grSU(3)$ representations $[n+1,1]$ with R-charges $y_0 = \frac{n}{3}$ and dimensions $\Delta_0 = \abs{y_0} + 1 = \frac{n+3}{3}$ for $n=0,1,2,\ldots$. When $n=0$, this is in fact the conserved current multiplet $\mathcal{J}^{(0)}_A{}^B$ corresponding to the $\grSU(3)$ global symmetry of the CFT. This superfield satisfies the constraint $D^2 \mathcal{J}^{(0)}_A{}^B = \bar{D}^2 \mathcal{J}^{(0)}_A{}^B = 0$. Its highest spin component is the bosonic current
\be
  J_{\mu A}^{(0)B} = \bar{Z}_A \!\stackrel{\leftrightarrow}{\partial}_\mu\! Z^B - \frac{1}{3}\delta_A^B \bar{Z}_C \!\stackrel{\leftrightarrow}{\partial}_\mu\! Z^C \; .
\ee
It has the protected classical dimension of $2$.
For higher $n$ we expect the series to continue as
\be
  \mathcal{J}^{(n)A_1 \ldots A_{n+1}}_{A_0} \sim \mathcal{J}_{A_0}^{(0)(A_1} \ZZ^{A_2} \cdots \ZZ^{A_{n+1})} \qquad \mbox{for $n=1,2,3,\ldots$} \; ,
\ee
where we still have to deal with the gauge indices. For $n\ge1$ these operators satisfy only the constraint $\bar{D}^2 \mathcal{J}^{(n)} = 0$.

To make these operators gauge invariant, we need $\lrfloor{\frac{n}{2}}$ monopole operators with two units of flux, $e^{-2\tau}$, and in case $n=\mbox{odd}$ another one with one unit of flux, $e^{-\tau}$. 
Since the latter ones do not exist for $k=2$, there are no operators for odd $n$, just as the corresponding SUGRA mode is projected out by the $k=2$ orbifold. The $e^{-2\tau}$ operators are inserted on every other $\ZZ$ just as for the hypermultiplet and summed over all possible permutations. One typical permutation is for example,
\be
  \tr \mathcal{J}_{A_0}^{(0)(A_1} \ZZ^{A_2} e^{-2\tau} \ZZ^{A_3} \cdots \ZZ^{A_{n}} e^{-2\tau} \ZZ^{ A_{n+1})} \; .
\ee

\section*{Acknowledgments}
We are very grateful to Marcus Benna and Mikael Smedb\" ack for their valuable input into various aspects of this paper. We are also grateful to Daniel Jafferis, Anton Kapustin and Juan Maldacena for very helpful discussions. This research is supported in part by the National Science Foundation Grant No.~PHY-0756966.

\appendix

\section{$\superN=2$ supermultiplets}
\label{sec:supermultiplets}

In the main text we have used the knowledge of the structure of $\grOsp(2|4)$ supermultiplets to constrain the spectrum of gravity states on the `stretched and squashed' seven sphere. These supermultiplets have been worked out in the context of general $\superN=2$ compactifications in \cite{Ceresole:1984hr} (see also \cite{Gualtieri:1999tu}). The short multiplets and their gauge theory interpretation in a general AdS$_4$/CFT$_3$ context were discussed in \cite{Fabbri:1999ay}. For the convenience of the reader we list the multiplets relevant to our discussion in this appendix.

The bosonic subgroup of $\grOsp(2|4)$ is $\grSO(3,2)\times\grSO(2)$. The $\grSO(3,2)$ part is the conformal group in 2+1 dimensions or, equivalently, the isometry group of $\AdS_4$. Unitary, positive energy representations of $\grSO(3,2)$ are labeled by spin $s$ and energy $E$ \cite{Heidenreich:1982rz}. The $\grSO(2)$ part is the R-symmetry and the representation label is the hypercharge $y$. An $\superN=2$ supermultiplet is a set of $\grSO(3,2)\times\grSO(2)$ representations which is obtained by acting with the fermionic raising operators of $\grOsp(2|4)$ onto a chosen $\grSO(3,2)\times\grSO(2)$ with labels $(s_0,E_0,y_0)$, the so-called lowest bosonic submultiplet.

The total number of bosonic submultiplets within one $\grOsp(2|4)$ representation depends on the relationships between the labels $(s_0,E_0,y_0)$:
\begin{itemize}
 \item Long multiplets for $E_0 > \abs{y_0} + s_0 + 1$: \\
       long graviton ($s_0=1$), long gravitino ($s_0=\half$), long vector ($s_0=0$),
 \item Short multiplets `I' for $E_0 = \abs{y_0} + s_0 + 1$: \\
       short graviton ($s_0=1$), short gravitino ($s_0=\half$), short vector ($s_0=0$),
 \item Short multiplets `II' for $E_0 = \abs{y_0} \ge \half$: \\
       hypermultiplet ($s_0=0$),
 \item Ultrashort multiplets for $E_0 = s_0+1$, $y_0=0$: \\
       massless graviton ($s_0=1$), massless vector ($s_0=0$).
\end{itemize}
Note that there is no massless gravitino as its presence would enhance the supersymmetry to $\superN > 2$.

\begin{table}[H]
\begin{center}
\begin{tabular}{|l|c|c|c|c|}
\hline
Spin & $2$ & $\frac{3}{2}$ & $\frac{3}{2}$ & $1$  \\  \hline
Energy & $3$ & $\frac{5}{2}$ & $\frac{5}{2}$ & $2$  \\ \hline
R-charge & $0$ & $+1$ & $-1$ & $0$ \\ \hline
\end{tabular}
\caption{\textbf{\mathversion{bold} $\superN=2$ massless graviton multiplet (MGRAV).}}
\label{tab:massless-graviton}
\end{center}
\end{table}

\begin{table}[H]
\begin{center}
\begin{tabular}{|l|c|c|c|c|c|c|c|c|}
\hline
Spin & $2$ & $\frac{3}{2}$ & $\frac{3}{2}$ & $\frac{3}{2}$ & $1$ &  $1$ & $1$ & $\frac{1}{2}$ \\ \hline
Energy & $E_0+1$ & $E_0+\frac{3}{2}$ & $E_0+\half$ & $E_0+\half$ & $E_0+1$ &  $E_0+1$ &  $E_0$ & $E_0+\half$ \\ \hline
R-charge & $y_0$ & $y_0\mp1$ & $y_0+1$ & $y_0-1$& $y_0\mp2$ & $y_0$ & $y_0$ & $y_0\mp1$ \\ \hline
\end{tabular}
\caption{\textbf{\mathversion{bold} $\superN=2$ short graviton multiplet (SGRAV).} $E_0 = \abs{y_0} + 2$}
\label{tab:short-graviton}
\end{center}
\end{table}

\begin{table}[H]
\begin{center}
\begin{tabular}{|l|c|c|c|c|c|c|c|c|c|c|c|}
\hline
Spin     & $2$     & $\frac{3}{2}$     & $\frac{3}{2}$     & $\frac{3}{2}$ & $\frac{3}{2}$ & $1$     & $1$     & $1$     \\ \hline
Energy   & $E_0+1$ & $E_0+\frac{3}{2}$ & $E_0+\frac{3}{2}$ & $E_0+\half$   & $E_0+\half$   & $E_0+2$ & $E_0+1$ & $E_0+1$ \\ \hline
R-charge & $y_0$   & $y_0-1$           & $y_0+1$           & $y_0-1$       & $y_0+1$       & $y_0$   & $y_0-2$ & $y_0+2$ \\ \hline \hline
Spin     & $1$     & $1$     & $1$   & $\frac{1}{2}$     & $\frac{1}{2}$     & $\frac{1}{2}$ & $\frac{1}{2}$ & $0$     \\ \hline
Energy   & $E_0+1$ & $E_0+1$ & $E_0$ & $E_0+\frac{3}{2}$ & $E_0+\frac{3}{2}$ & $E_0+\half$   & $E_0+\half$   & $E_0+1$ \\ \hline
R-charge & $y_0$   & $y_0$   & $y_0$ & $y_0-1$           & $y_0+1$           & $y_0-1$       & $y_0+1$       & $y_0$   \\ \hline
\end{tabular}
\caption{\textbf{\mathversion{bold} $\superN=2$ long graviton multiplet (LGRAV).}}
\label{tab:long-graviton}
\end{center}
\end{table}

\begin{table}[H]
\begin{center}
\begin{tabular}{|l|c|c|c|c|c|c|c|c|}
\hline
Spin & $\frac{3}{2}$ & 1& 1& 1&  $\frac{1}{2}$ & $\frac{1}{2}$ & $\frac{1}{2}$ & $0$  \\  \hline
Energy & $E_0+1$ & $E_0+\half$ & $E_0+\half$ & $E_0+\frac{3}{2}$ &  $E_0+1$ &  $E_0+1$ &  $E_0$ & $E_0+\half$ \\ \hline
R-charge & $y_0$ & $y_0-1$ & $y_0+1$& $y_0\mp1$ & $y_0\mp2$ & $y_0$ & $y_0$ & $y_0\mp1$ \\ \hline
\end{tabular}
\caption{\textbf{\mathversion{bold} $\superN=2$ short gravitino multiplet (SGINO).} $E_0 = \abs{y_0} + \frac{3}{2}$}
\label{tab:short-gravitino}
\end{center}
\end{table}

\begin{table}[H]
\begin{center}
\begin{tabular}{|l|c|c|c|c|c|c|c|c|c|c|c|}
\hline
Spin     & $\frac{3}{2}$ & $1$               & $1$               & $1$           & $1$           & $\half$ & $\half$ & $\half$ \\ \hline
Energy   & $E_0+1$       & $E_0+\frac{3}{2}$ & $E_0+\frac{3}{2}$ & $E_0+\half$   & $E_0+\half$   & $E_0+2$ & $E_0+1$ & $E_0+1$ \\ \hline
R-charge & $y_0$         & $y_0-1$           & $y_0+1$           & $y_0-1$       & $y_0+1$       & $y_0$   & $y_0-2$ & $y_0$ \\ \hline \cline{1-8}
Spin     & $\half$ & $\half$ & $\half$ & $0$               & $0$               & $0$         & $0$         \\ \cline{1-8}
Energy   & $E_0+1$ & $E_0+1$ & $E_0$   & $E_0+\frac{3}{2}$ & $E_0+\frac{3}{2}$ & $E_0+\half$ & $E_0+\half$ \\ \cline{1-8}
R-charge & $y_0+2$ & $y_0$   & $y_0$   & $y_0-1$           & $y_0+1$           & $y_0-1$     & $y_0+1$     \\ \cline{1-8}
\end{tabular}
\caption{\textbf{\mathversion{bold} $\superN=2$ long gravitino multiplet (LGINO).}}
\label{tab:long-gravitino}
\end{center}
\end{table}

\begin{table}[H]
\begin{center}
\begin{tabular}{|l|c|c|c|c|c|}
\hline
Spin & $1$ & $\frac{1}{2}$ & $\frac{1}{2}$ & $0$ & $0$ \\  \hline
Energy & $2$ & $\frac{3}{2}$ & $\frac{3}{2}$ & $2$ & $1$ \\ \hline
R-charge & $0$ & $+1$ & $-1$ & $0$ & $0$ \\ \hline
\end{tabular}
\caption{\textbf{\mathversion{bold} $\superN=2$ massless vector multiplet (MVEC).}}
\label{tab:massless-vector}
\end{center}
\end{table}

\begin{table}[H]
\begin{center}
\begin{tabular}{|l|c|c|c|c|c|c|c|c|c|c|}
\hline
Spin & $1$ & $\frac{1}{2}$ & $\frac{1}{2}$ & $\frac{1}{2}$ & $0$ & $0$ & $0$ \\  \hline
Energy & $E_0 + 1$ & $E_0 + \frac{3}{2}$ & $E_0 + \half$ & $E_0 + \half$ & $E_0 + 1$ & $E_0 + 1$ & $E_0$ \\ \hline
R-charge & $y_0$ & $y_0\mp1$ & $y_0-1$ & $y_0+1$ & $y_0\mp2$ & $y_0$ & $y_0$ \\ \hline
\end{tabular}
\caption{\textbf{\mathversion{bold} $\superN=2$ short vector multiplet (SVEC).} $E_0 = \abs{y_0} + 1$}
\label{tab:short-vector}
\end{center}
\end{table}

\begin{table}[H]
\begin{center}
\begin{tabular}{|l|c|c|c|c|c|c|c|c|c|c|}
\hline
Spin     & $1$     & $\half$           & $\half$           & $\half$           & $\half$           \\ \hline
Energy   & $E_0+1$ & $E_0+\frac{3}{2}$ & $E_0+\frac{3}{2}$ & $E_0+\frac{1}{2}$ & $E_0+\frac{1}{2}$ \\ \hline
R-charge & $y_0$   & $y_0-1$           & $y_0+1$           & $y_0-1$           & $y_0+1$           \\ \hline \hline
Spin     & $0$     & $0$       & $0$     & $0$     & $0$   \\ \hline
Energy   & $E_0+2$ & $E_0+1$   & $E_0+1$ & $E_0+1$ & $E_0$ \\ \hline
R-charge & $y_0$   & $y_0-2$   & $y_0$   & $y_0+2$ & $y_0$ \\ \hline
\end{tabular}
\caption{\textbf{\mathversion{bold} $\superN=2$ long vector multiplet (LVEC).}}
\label{tab:massive-vector}
\end{center}
\end{table}

\begin{table}[H]
\begin{center}
\begin{tabular}{|l|c|c|c|c|}
\hline
Spin &  $\frac{1}{2}$ & $0$ & $0$  \\  \hline
Energy & $E_0 + \half$ & $E_0$ & $E_0 + 1$ \\ \hline
R-charge & $y_0\mp1$ & $y_0$ & $ y_0\mp2$ \\ \hline
\end{tabular}
\caption{\textbf{\mathversion{bold} $\superN=2$ hyper multiplet (HYP).} $E_0 = \abs{y_0}$}
\label{tab:massive-hyper}
\end{center}
\end{table}

\section{Choices of dressing for the lowest hypermultiplet }
\label{app:dressing}

In this appendix we make a curious observation which relates the operator dimensions of the fields in the hypermultiplet in Scenario I to the ones in Scenario II at the massless level originally studied in \cite{Nicolai:1985hs}. Recall that in Scenario I the hypermultiplet contains scalar operators of dimension $\frac{2}{3}$ and $\frac{5}{3}$, and a fermionic operator of dimension $\frac{7}{6}$; in Scenario II it contains scalar operators of dimension $\frac{4}{3}$ and $\frac{7}{3}$, and a fermionic operator of dimension $\frac{11}{6}$. We show that the three mass-squared values of the fields comprising these hypermultiplets are the same for the two scenarios, but they differ only in the choice of the branches in the formulae for the dimension. For scalars in $AdS_4$ the corresponding operators have dimensions
\be
\Delta_\pm = \frac{3}{2} \pm \sqrt{ \frac{9}{4}+ m^2 }
\ee
and both choices are allowed \cite{Klebanov:1999tb} for $-\frac{9}{4}< m^2 < -\frac{5}{4}$. For a scalar of $m^2=-\frac{14}{9}$, we find
that $\Delta_-=\frac{2}{3}$ giving the ground state of the Scenario I multiplet, while $\Delta_+=\frac{7}{3}$ corresponding to the second scalar in the Scenario II multiplet. Similarly, for $m^2=-\frac{20}{9}$, $\Delta_-=\frac{4}{3}$ giving the ground state scalar in Scenario II, while $\Delta_+=\frac{5}{3}$ corresponding to the second scalar in Scenario I.
For the fermionic operators the correct formula is \cite{Contino:2004vy}
\be
\Delta_f = 1+\abs{m+\half} \; .
\ee
We find that with $m^2=\frac{1}{9}$ the two choices of sign, $m=\pm \frac{1}{3}$, reproduce dimensions $\frac{7}{6}$ and $\frac{11}{6}$. Thus, for this part of the spectrum the distinction between the two scenarios does not concern the $m^2$ spectrum in $AdS_4$ but only the boundary conditions. However, we note that this relationship does not persist to higher levels where completely different values of $m^2$ occur in the two scenarios.

\section{Supermultiplets at higher levels}
\label{sec:higher-levels}

In this appendix, we list the $\superN=2$ supermultiplets of gravity states at the first few Kaluza-Klein levels $n$. We group them according to the $\grSU(3)$ representations $[a,b]$ under which they transform. One observes that at level $n$ exactly those $\grSU(3)$ representations occur which satisfy $a+b\le n+2$. Furthermore, the supermultiplets with representation $[b,a]$ are conjugate to the ones in the representation $[a,b]$ in the sense that their R-charge is negated.

In the first subsection of this appendix we present the spectrum following from the embedding of $\grSU(3)\times\grU(1)_R$ into $\grSO(8)$ which yields agreement with the gauge theory (Scenario I). For comparison we also exhibit the first few levels of the spectrum resulting form Scenario II in the second subsection. The acronyms as MGRAV, SGINO, etc. refer to the $\superN=2$ supermultiplets defined in the tables \ref{tab:massless-graviton} to \ref{tab:massive-hyper} in \appref{sec:supermultiplets}. The numbers following the acronyms specify the R-charges of the supermultiplets of this kind.

Since parity is broken, there are some ambiguities for grouping the states into supermultiplets. For certain ranges of R-charges one finds  $\mathrm{SVEC}_y \cup \mathrm{HYP}_{y+2} = \mathrm{LVEC}_y$ and $\mathrm{SGRAV}_y \cup \mathrm{SGINO}_{y+1} = \mathrm{LGRAV}_y$. In these cases we have noted the long multiplets in the tables below. These ambiguities can only be resolved by an explicit KK reduction, but in any case they do not affect the four series of short operators which we are mainly interested in.

\subsection{Scenario I}

\begin{table}[H]
\begin{center}
{\footnotesize
\begin{tabular}{|p{20mm}|p{20mm}|p{20mm}|} \hline
$[0,0]$ & $[0,1]$ & $[0,2]$ \\
MGRAV $0$ & SGINO $-\frac{1}{3}$ & HYP $-\frac{2}{3}$ \\
LVEC $0$ & & \\ \hline
$[1,0]$ & $[1,1]$ \\
SGINO $+\frac{1}{3}$ & MVEC $0$ \\ \cline{1-2}
$[2,0]$ \\
HYP $+\frac{2}{3}$ \\ \cline{1-1}
\end{tabular}
}
\caption{\textbf{\mathversion{bold} Multiplets of IR theory at level $n=0$.}}
\label{tab:n0}
\end{center}
\end{table}

\begin{table}[H]
\begin{center}
{\footnotesize
\begin{tabular}{|p{30mm}|p{30mm}|p{30mm}|p{30mm}|} \hline
$[0,0]$ & $[0,1]$ & $[0,2]$ & $[0,3]$ \\
SGRAV $+1$, $-1$ & LGRAV $-\frac{1}{3}$ & SGINO $-\frac{2}{3}$ & HYP $-1$ \\
LVEC  $+1$, $-1$ & LGINO $+\frac{2}{3}$ & LVEC  $+\frac{1}{3}$   &  \\
                 & LVEC  $-\frac{1}{3}$ &                      &  \\ \hline
$[1,0]$ & $[1,1]$ & $[1,2]$ \\
LGRAV $+\frac{1}{3}$ & LGINO $0$ & SVEC $-\frac{1}{3}$ \\
LGINO $-\frac{2}{3}$ &           &                    \\
LVEC  $+\frac{1}{3}$ &           &                    \\ \cline{1-3}
$[2,0]$ & $[2,1]$ \\
SGINO $+\frac{2}{3}$ & SVEC $+\frac{1}{3}$ \\
LVEC  $-\frac{1}{3}$ &                    \\ \cline{1-2}
$[3,0]$ \\
HYP $+1$ \\ \cline{1-1}
\end{tabular}
}
\caption{\textbf{\mathversion{bold} Multiplets of IR theory at level $n=1$.}}
\label{tab:n1}
\end{center}
\end{table}

\begin{table}[H]
\begin{center}
\begin{sideways}
{\footnotesize
\begin{tabular}{|l|l|l|l|l|} \hline
$[0,0]$ & $[0,1]$ & $[0,2]$ & $[0,3]$ & $[0,4]$ \\
LGRAV $0$ & LGRAV $-\frac{4}{3}$, $+\frac{2}{3}$ & LGRAV $-\frac{2}{3}$ & SGINO $-1$ & HYP $-\frac{4}{3}$ \\
SGRAV $-2$, $+2$ & LGINO $-\frac{1}{3}$, $-\frac{1}{3}$, $+\frac{5}{3}$ & LGINO $+\frac{1}{3}$ & LVEC $0$ & \\
LVEC  $-2$, $0$, $+2$ & LVEC $-\frac{4}{3}$, $+\frac{2}{3}$ & LVEC $-\frac{2}{3}$, $-\frac{2}{3}$, $+\frac{4}{3}$ & & \\ \hline
$[1,0]$ & $[1,1]$ & $[1,2]$ & $[1,3]$ \\
LGRAV $-\frac{2}{3}$, $+\frac{4}{3}$ & LGRAV $0$ & LGINO $-\frac{1}{3}$, $-\frac{1}{3}$ & SVEC $-\frac{2}{3}$ \\
LGINO $-\frac{5}{3}$, $+\frac{1}{3}$, $+\frac{1}{3}$ & LGINO $-1$, $-1$, $+1$, $+1$ & LVEC $+\frac{2}{3}$ & \\
LVEC  $-\frac{2}{3}$, $+\frac{4}{3}$ & LVEC $0$, $0$ & & \\ \cline{1-4}
$[2,0]$ & $[2,1]$ & $[2,2]$ \\
LGRAV $+\frac{2}{3}$ & LGINO $+\frac{1}{3}$, $+\frac{1}{3}$ & LVEC $0$ \\
LGINO $-\frac{1}{3}$ & LVEC $-\frac{2}{3}$ & \\
LVEC  $-\frac{4}{3}$, $+\frac{2}{3}$, $+\frac{2}{3}$ & & \\ \cline{1-3}
$[3,0]$ & $[3,1]$ \\
SGINO $+1$ & SVEC $\frac{2}{3}$ \\
LVEC $0$   & \\ \cline{1-2}
$[4,0]$ \\
HYP $+\frac{4}{3}$ \\ \cline{1-1}
\end{tabular}
}
\end{sideways}
\caption{\textbf{\mathversion{bold} Multiplets of IR theory at level $n=2$.}}
\label{tab:n2}
\end{center}
\end{table}

\begin{table}[H]
\begin{center}
\begin{sideways}
{\footnotesize
\begin{tabular}{|l|l|l|l|l|l|} \hline
$[0,0]$ & $[0,1]$ & $[0,2]$ & $[0,3]$ & $[0,4]$ & $[0,5]$ \\
LGRAV $-1$, $+1$ & conj. to $[1,0]$ & conj. to $[2,0]$ & conj. to $[3,0]$ & conj. to $[4,0]$ & conj. to $[5,0]$ \\
SGRAV $-3$, $+3$ &       &       &       &       &       \\
LVEC $-3$, $-1$, $-1$, $+1$, $+1$, $+3$ & & & & &        \\ \hline
$[1,0]$ & $[1,1]$ & $[1,2]$ & $[1,3]$ & $[1,4]$ \\
LGRAV $-\frac{5}{3}$, $+\frac{1}{3}$, $+\frac{7}{3}$ & LGRAV $-1$, $+1$ & conj. to $[2,1]$ & conj. to $[3,1]$ & conj. to $[4,1]$ \\
LGINO $-\frac{8}{3}$, $-\frac{2}{3}$, $-\frac{2}{3}$, $+\frac{4}{3}$, $+\frac{4}{3}$ & LGINO $-2$, $-2$, $0$, $0$, $0$, $0$, $+2$, $+2$ & & & \\
LVEC $-\frac{5}{3}$, $-\frac{1}{3}$, $-\frac{1}{3}$, $+\frac{7}{3}$ & LVEC $-1$, $-1$, $+1$, $+1$ & & & \\ \cline{1-5}
$[2,0]$ & $[2,1]$ & $[2,2]$ & $[2,3]$ \\
LGRAV $-\frac{1}{3}$, $+\frac{5}{3}$ & LGRAV $+\frac{1}{3}$ & LGINO $0$, $0$ & conj. to $[3,2]$ \\
LGINO $-\frac{4}{3}$, $+\frac{2}{3}$, $+\frac{2}{3}$ & LGINO $-\frac{2}{3}$, $-\frac{2}{3}$, $+\frac{4}{3}$, $+\frac{4}{3}$ & LVEC $-1$, $+1$ & \\
LVEC  $-\frac{7}{3}$, $-\frac{1}{3}$, $-\frac{1}{3}$, $+\frac{5}{3}$, $+\frac{5}{3}$ & LVEC $-\frac{5}{3}$, $+\frac{1}{3}$, $+\frac{1}{3}$, $+\frac{1}{3}$ & & \\ \cline{1-4}
$[3,0]$ & $[3,1]$ & $[3,2]$ \\
LGRAV $+1$ & LGINO $+\frac{2}{3}$, $+\frac{2}{3}$ & LVEC $+\frac{1}{3}$ \\
LGINO $0$ & LVEC $-\frac{1}{3}$ & \\
LVEC $-1$, $+1$, $+1$ & & \\ \cline{1-3}
$[4,0]$ & $[4,1]$ \\
SGINO $+\frac{4}{3}$ & SVEC $+1$ \\
LVEC $+\frac{1}{3}$ & \\ \cline{1-2}
$[5,0]$ \\
HYP $+\frac{5}{3}$ \\ \cline{1-1}
\end{tabular}
}
\end{sideways}
\caption{\textbf{\mathversion{bold} Multiplets of IR theory at level $n=3$.}}
\label{tab:n3}
\end{center}
\end{table}

\newpage
\subsection{Scenario II}

\begin{table}[H]
\begin{center}
{\footnotesize
\begin{tabular}{|p{20mm}|p{20mm}|p{20mm}|} \hline
$[0,0]$ & $[0,1]$ & $[0,2]$ \\
MGRAV $0$ & SGINO $-\frac{1}{3}$ & HYP $\frac{4}{3}$ \\
LVEC $0$ & & \\ \hline
$[1,0]$ & $[1,1]$ \\
SGINO $+\frac{1}{3}$ & MVEC $0$ \\ \cline{1-2}
$[2,0]$ \\
HYP $-\frac{4}{3}$ \\ \cline{1-1}
\end{tabular}
}
\caption{\textbf{\mathversion{bold} Multiplets of Scenario II IR theory at level $n=0$.}}
\label{tab:n0-Warner}
\end{center}
\end{table}

\begin{table}[H]
\begin{center}
{\footnotesize
\begin{tabular}{|l|l|l|l|} \hline
$[0,0]$ & $[0,1]$ & $[0,2]$ & $[0,3]$ \\
SGRAV $+0$, $-0$ & conj. to $[1,0]$ & conj. to $[2,0]$ & conj. to $[3,0]$ \\
LVEC $0$, $0$ &  &   &  \\ \hline
$[1,0]$ & $[1,1]$ & $[1,2]$ \\
SGRAV $-\frac{2}{3}$ & LGINO $0$ & conj. to $[2,1]$ \\
LGINO $+\frac{1}{3}$ &           &                    \\
SGINO $+\frac{1}{3}$ &           &                    \\
LVEC  $-\frac{2}{3}$ &           &                    \\ \cline{1-3}
$[2,0]$ & $[2,1]$ \\
SGINO $-\frac{1}{3}$ & SVEC $-\frac{2}{3}$ \\
SVEC  $-\frac{4}{3}$ &                    \\
HYP   $-\frac{4}{3}$ &                    \\ \cline{1-2}
$[3,0]$ \\
HYP $-2$ \\ \cline{1-1}
\end{tabular}
}
\caption{\textbf{\mathversion{bold} Multiplets of Scenario II IR theory at level $n=1$.}}
\label{tab:n1-Warner}
\end{center}
\end{table}

\begin{table}[H]
\begin{center}
\begin{sideways}
{\footnotesize
\begin{tabular}{|l|l|l|l|l|} \hline
$[0,0]$ & $[0,1]$ & $[0,2]$ & $[0,3]$ & $[0,4]$ \\
LGRAV $0$ & conj. to $[1,0]$ & conj. to $[2,0]$ & conj. to $[3,0]$ & conj. to $[4,0]$ \\
SGRAV $-0$, $+0$ & & &  & \\
LVEC  $0$, $0$, $0$, $0$ & & & & \\ \hline
$[1,0]$ & $[1,1]$ & $[1,2]$ & $[1,3]$ \\
LGRAV $-\frac{2}{3}$ & LGRAV $0$ & conj. to $[2,1]$ & conj. to $[3,1]$ \\
SGRAV $-\frac{2}{3}$ & LGINO $-1$, $+1$ & &  \\
LGINO $+\frac{1}{3}$, $+\frac{1}{3}$, $+\frac{1}{3}$ & SGINO $-1$, $+1$ &  & \\
SGINO $+\frac{1}{3}$ & SVEC $-0$, $-0$, $+0$, $+0$ & & \\
LVEC  $-\frac{2}{3}$, $-\frac{2}{3}$ & HYP $-2$, $+2$ & & \\ \cline{1-4}
$[2,0]$ & $[2,1]$ & $[2,2]$ \\
SGRAV $-\frac{4}{3}$ & LGINO $+\frac{1}{3}$ & LVEC $0$ \\
LGINO $-\frac{1}{3}$, $-\frac{1}{3}$ & SGINO $-\frac{5}{3}$ & \\
LVEC  $-\frac{4}{3}$, $+\frac{2}{3}$ & LVEC $-\frac{2}{3}$ & \\
HYP   $-\frac{4}{3}$ & SVEC $-\frac{2}{3}$ & \\ \cline{1-3}
$[3,0]$ & $[3,1]$ \\
SGINO $-1$ & SVEC $-\frac{4}{3}$ \\
SVEC $-2$   & \\
HYP $-2$   & \\   \cline{1-2}
$[4,0]$ \\
HYP $-\frac{8}{3}$ \\ \cline{1-1}
\end{tabular}
}
\end{sideways}
\caption{\textbf{\mathversion{bold} Multiplets of Scenario II IR theory at level $n=2$.}}
\label{tab:n2-Warner}
\end{center}
\end{table}

\section{Monopole Operators}
\label{app:monopole}

The monopole (or 't Hooft) operators in $2+1$ dimensions can be viewed as changing the boundary conditions for fields in the path integral in a way that produces some specified magnetic flux through an $S^2$ around some point $x$. Hence these can also be called monopole creation operators \cite{Kap2002} and are local.

We can classify the flux of magnetic monopoles in a 3d gauge theory using the scheme in \cite{GNO}\footnote{The theories of interest in \cite{GNO} were 4d gauge theories but the monopoles were time-independent objects identical to what we wish to insert in our 3d gauge theory.}. We take the singularity to be of the form,
\be
F \sim *  d\left(\frac{1}{|x|}\right) M
\ee
where $M$ is some generator of the gauge group $G$. The generalized Dirac quantization condition is,
\be
e^{2 \pi i M} = 1
\ee
By conjugation, $M$ can be brought to the form $\beta^a G_a$ where $G_a$ are the Cartan generators of $G$.

When there is a Chern-Simons term with level $k$, such monopoles transform in a representation of $G$. For example, consider an abelian theory on $S^2 \times R$ (i.e in the radial quantization picture) with the Chern-Simons term $k \int A \wedge dA$. With $n$ units of flux through the $S^2$, we can integrate the Chern-Simons term over $S^2$ to obtain $k n \int A_0 dt$ which is a coupling to a particle of charge $k n$.

In general, a monopole with flux $\beta^a$ transforms in the representation of $G$ with highest weight state given by $k \beta^a$. Let us illustrate this in the case of $\grU(N)$. The quantization condition is solved (up to conjugation) by $M$ in the form of a diagonal matrix $\diag(m_1,m_2,\ldots,m_N)$ with $m_1 \geq m_2 \geq \ldots \geq m_N$ all being integers (cf. \cite{KW}). Such a monopole would transform in a representation of $\grU(N)$ with the highest weight state given by $(k m_1,k m_2,\ldots,k m_N)$. In the notation of \cite{KW}, this corresponds to a Young tableaux with rows of length $k m_1, k m_2, \ldots , k m_N$. We note that since we are interested in representations of $\grU(N)$ and not $\grSU(N)$, we must keep track of columns of length $N$ since they give the charge under the central $\grU(1)$ subgroup of $\grU(N)$.

Turning our attention to the $\grU(N)\times \grU(N)$ gauge theory of interest, we will be interested in monopole operators of the form $(e^{n \tau})^{a_1 \ldots a_n}_{\ha_1 \ldots \ha_n}$ which transform in conjugate representations of the two gauge groups. Hence we give the choice of flux $M$ in the first group alone. The conjugate representation is understood to be chosen in the other $\grU(N)$.

\subsubsection*{$k=1$}
The basic monopole operator for $k=1$ transforms in the bi-fundamental representation with the simplest choice of flux,
\be
M = \diag(1,0,0,\ldots,0) \quad (e^{\tau})^{a}_{\ha}
\ee
It can be used to render operators with odd powers of $\ZZ$ gauge-invariant \cite{Aharony:2008ug}. For operators with two indices in each group, we have the following choices for the flux giving symmetric and anti-symmetric operators,
\begin{align}
M & = \diag (2,0,0,\ldots,0) & (e^{2 \tau})^{ab}_{\ha \hb} & =   (e^{2 \tau})^{ba}_{\ha \hb} = (e^{2 \tau})^{ba}_{\hb \ha} \; , \\
M & = \diag(1,1,0,\ldots,0)  & (e^{2 \tau})^{ab}_{\ha \hb} & = - (e^{2 \tau})^{ba}_{\ha \hb} = (e^{2 \tau})^{ba}_{\hb \ha} \; .
\end{align}
The symmetric operators were used in \cite{Aharony:2008ug} while the anti-symmetric operators are important in writing down the mass deformation discussed in this paper. Note that both choices are symmetric under the simultaneous interchange of both sets of indices.

When $N=2$, the anti-symmetric operator can also be viewed as an abelian monopole operator creating flux for $\grU(1)_{\mathrm{diag}}$ of $\grU(2) \times \grU(2)$ which hence carries $\grU(1)_b$ charge due to the Chern-Simons term of ABJM theory as explained in \cite{Aharony:2008ug,Lambert:2008et,Distler:2008mk}. Hence it was denoted $(e^{2\tau})^{ab}_{\ha\hb} = T^2 \levi^{ab} \levi_{\ha \hb} $ in this paper where $T^2$ is the abelian operator with two units of $\grU(1)_b$ charge and creates two units of flux for $\grU(1)_{\mathrm{diag}}$.

\subsubsection*{$k=2$}
When $k=2$, one cannot construct a monopole operator with the indices $(e^{\tau})^{a}_{\ha}$. The smallest choice of flux $M = \diag(1,0,\ldots,0)$, after multiplying by $k=2$, already corresponds to an operator with two pairs of indices, $(e^{2 \tau})^{ab}_{\ha \hb}$ symmetric in upper and lower indices separately,
\begin{align}
k M & = \diag(2,0,0,\ldots,0) & (e^{2 \tau})^{ab}_{\ha \hb} & = (e^{2 \tau})^{ba}_{\ha \hb} = (e^{2 \tau})^{ba}_{\hb \ha} \; .
\end{align}

Trying to form an anti-symmetric operator with 2 indices fails since we would need $k M = \diag(1,1,0,\ldots,0)$ but such a $M$ would not obey the Dirac quantization above for general $N$. However, when $N=2$, we can effectively create an anti-symmetric operator by using an abelian monopole operator charged under $\grU(1)_b$ as in the $k=1$ case. Such an operator can again be written as
\be
  (e^{2\tau})^{ab}_{\ha\hb} = T^2 \levi^{ab} \levi_{\ha \hb} \; .
\ee
$T^2$ is again an abelian monopole operator with two units of $\grU(1)_b$ charge but since $k=2$, this requires turning on only one unit of flux for $\grU(1)_{\mathrm{diag}}$ unlike in the $k=1$ case above. Formally, we can assign such an operator the flux $\diag(\frac{1}{2},\frac{1}{2})$. This satisfies the fractional quantization condition $e^{2 \pi i M} = - 1 \in Z(\grSU(2))$.\footnote{An important modification of the quantization condition occurs when the gauge group has a non-trivial center under which all the matter transform trivially. The gauge group is then effectively $G/Z(G)$ where $Z(G)$ is the center and the quantization condition is then $ e^{2 \pi i M} \in Z(G)$. For example, in a $\grSU(N)$ gauge theory with adjoint matter, the $\Integers_N$ subgroup decouples. When $N=2$, this allows for example $M = \diag(\frac{1}{2},\frac{1}{2})$ when the gauge group is taken to be $\grSU(2)/\Integers_2$ since $e^{2 \pi i M}= -1$. This is in addition those $M$ satisfying $e^{2 \pi i M} = 1$ allowed when the gauge group is $\grSU(2)$.}

\newpage
\bibliographystyle{nb}
\bibliography{AdS4CFT3}

\end{document}